\documentclass[journal,10pt]{IEEEtran}
\normalsize

\usepackage{cite}
\usepackage{graphicx}
\usepackage[cmex10]{amsmath}
\usepackage{amssymb}
\usepackage{booktabs}
\usepackage{threeparttable}
\usepackage{algorithm}
\usepackage{algpseudocode}
\usepackage{amsfonts}
\usepackage{xcolor}
\usepackage{color}

\usepackage{tabularx} %

\usepackage{caption}

\usepackage{enumitem}
\usepackage{framed}

\usepackage{dblfloatfix} %

\usepackage{soul}

\usepackage{cancel} %

\usepackage{verbatim}%

\usepackage{makecell}

\usepackage{empheq} %

\usepackage{underoverlap} %

\usepackage{hyperref}
\hypersetup{pdftex,colorlinks=true,allcolors=black}

\usepackage{calc}

\usepackage{textcomp}
\def\BibTeX{{\rm B\kern-.05em{\sc i\kern-.025em b}\kern-.08em
		T\kern-.1667em\lower.7ex\hbox{E}\kern-.125emX}}

\newtheorem{remark}{\bf  Remark}

\newtheorem{proposition}{\bf Proposition}

\newtheorem{lemma}{\bf  Lemma}

\newcommand{\ma}{\mathsf{g}}

\newcommand{\mb}{\mathsf{b}}

\newcommand{\me}{\mathsf{e}}

\newcommand{\mC}{\mathsf{C}}

\newcommand{\mP}{\mathsf{P}}
\newcommand{\mT}{\mathsf{T}}

\newcommand{\mk}{\mathsf{k}}

\newcommand{\ml}{\mathsf{l}}
\newcommand{\mj}{\mathsf{j}}
\newcommand{\ts}{T_{\mathrm{s}}}

\newcommand{\mX}{\mathsf{X}}

\newcommand{\scpi}{\tilde{s}_{\mathrm{CP}}[i]}
\newcommand{\srcpi}{\tilde{s}_{\mathrm{RCP}}[i]}

\newcommand{\dso}{DFT-S-OFDM}
\newcommand{\rcpotfs}{RCP-OTFS}

\newcommand{\ccos}{C-COS}

\newcommand{\myDef}{\overset{\Delta}{=}}
\newcommand{\myEqualOverset}[1]{\overset{#1}{=}}
\newcommand{\myApprOverset}[1]{\overset{#1}{\approx}}

\newcommand{\df}{\Delta_f}
\newcommand{\dnu}{\Delta_{\nu}}

\newcommand{\myFloor}[1]{\left\lfloor #1 \right\rfloor}
\newcommand{\myExp}[1]{\mathbb{E}\left\{#1\right\}}
\newcommand{\myVar}[1]{\mathbb{V}\left\{#1\right\}}

\newcommand{\cn}[1]{\mathcal{CN}\left(0,#1\right)}
\newcommand{\myCN}[1]{\mathcal{CN}\left(#1\right)}
\newcommand{\myN}[1]{\mathcal{N}\left(#1\right)}
\newcommand{\myCC}[1]{\mathbb{C}\left(#1\right)} %

\newcommand{\myModulo}[2]{\left\langle#1\right\rangle_{#2}}
\newcommand{\myDFT}[2]{\mathcal{Z}_{#1}^{#2}}
\newcommand{\mySinc}[2]{\mathcal{S}_{#1}\left( #2 \right)}

\newcommand{\mySpaceTwoMM}{\vspace{2mm}}

\begin{document}

\title{Integrating Low-Complexity and Flexible Sensing into Communication Systems
}

\author{
	Kai Wu,
	 J. Andrew Zhang,~\IEEEmembership{Senior Member,~IEEE}, %
	Xiaojing Huang,~\IEEEmembership{Senior Member,~IEEE}, and\\
	Y. Jay Guo,~\IEEEmembership{Fellow,~IEEE}
	\thanks{K. Wu, J. A. Zhang, X. Huang and Y. J. Guo are with the Global Big Data Technologies Centre, University of Technology Sydney, Sydney, NSW 2007, Australia (e-mail: kai.wu@uts.edu.au; andrew.zhang@uts.edu.au; xiaojing.huang@uts.edu.au;   jay.guo@uts.edu.au).}%
\vspace{-1cm}	
}

\maketitle

\begin{abstract}	
	Integrating sensing into standardized communication systems can potentially benefit many consumer applications that require both radio frequency functions. However, without an effective sensing method, such integration may not achieve the expected gains of cost and energy efficiency. 
	Existing sensing methods, which use communication payload signals, either have limited sensing performance or suffer from high complexity. 
	In this paper, we develop a novel and flexible sensing framework which has a complexity only dominated by a Fourier transform and also provides the flexibility in adapting for different sensing needs. 
	We propose to segment a whole block of echo signal evenly into sub-blocks; adjacent ones are allowed to overlap. 
	We design a virtual cyclic prefix (VCP) for each sub-block that 
	allows us to employ two common ways of removing communication data symbols and generate two types of range-Doppler maps (RDMs) for sensing. We perform a comprehensive analysis of the signal components in the RDMs, proving that their interference-plus-noise (IN) terms are approximately Gaussian distributed. The statistical properties of the distributions are derived, which leads to 
	the analytical comparisons between the two RDMs as well as between the prior and our sensing methods. 
	Moreover, the impact of the lengths of sub-block, VCP
	and overlapping signal on sensing performance is analyzed. Criteria for designing these lengths for better sensing performance are also provided. Extensive simulations validate the superiority of the proposed sensing framework over prior methods in terms of 
	signal-to-IN ratios in RDMs, detecting performance and flexibility.

\end{abstract}


\vspace{-5mm}

\section{Introduction}\label{sec: introduction}

Integrated sensing and communications (ISAC) 
has attracted extensive attention recently. By allowing sensing and communications
to share the same waveform, hardware and frequency spectrum etc., ISAC not only improves cost and energy efficiency but also helps alleviate the increasingly severe congestion of frequency spectrum~\cite{FanLiu_overview2020TCOM}. 
As popularly seen in the literature, ISAC designs can be sensing-centric (SC) \cite{DFRC_SP_Mag2019Amin_Aboutanios,DFRC_automotive2020SPmag,DFRC_radarCentric_mishra2019toward,Kai_overviewFHMIMO_DFRC2020AES}, communication-centric (CC) \cite{DFRC_802p11ad2018TVT_Kumari,DFRC_OpportunisticRadar_80211ad_2018TSP,DFRC_CommCentric_duggal2020doppler,DFRC_QixunZhang9162963,DFRC_dsss2011procIeee,DFRC_SC_OFDM} and joint-design \cite{DFRC_JointDesign_9354629,DFRC_JointDesign_9148935,DFRC_AngLi_liu2021cram}. While SC-centric (or CC-) adds communications (or sensing) into existing sensing (or communications) systems as a secondary function, a joint-design ISAC generally solves a holistically formulated optimization problem for a (sub-)optimal DF waveform \cite{Andrew_jcasOverview2021JSTSP,Kai_rahman2020enablingSurvey,DFRC_wei2021towards}. 
In this paper, we focus on CC-ISAC which can potentially expedite the market penetration of ISAC into consumer markets \cite{DFRC_4IoT_cui2021integrating}. 
Moreover, CC-ISAC considered here performs active sensing; c.f., passive sensing based on communication signals.

A widely studied issue of CC-ISAC is how to achieve satisfactory sensing performance based on standardized communication waveforms. 
OFDM and its variant waveforms are popular in CC-ISAC. 
Note that sensing here is similar to the conventional radar sensing, i.e., detecting targets and estimating their parameters (mostly location and velocity) from the echo signals.  
In \cite{DFRC_dsss2011procIeee}, orthogonal frequency-division multiplexing (OFDM) waveform-based sensing is developed. The method has been widely studied since its development; see \cite{OFDM_autonomousDriv2019microwaveMag,BinYang_ofdmSPmagazine} and their references. Thus, we call it the classical OFDM sensing (COS). In short, 
COS i) transforms a block of OFDM symbols into the frequency domain; ii) removes communication data symbols through a point-wise division; iii) takes a two-dimensional DFT to generate a so-called range-Doppler map (RDM); and iv) performs target detection and estimation using the RDM. 
COS will be further reviewed in Section \ref{subsec: reviewing classical ofdm sensing} using our signal model.

In \cite{DFRC_SC_OFDM}, a COS-like sensing method is developed for DFT-spread OFDM (\dso) waveforms. 
Different from OFDM, \dso~presents 
Gaussian randomness in the frequency domain. Hence, directly dividing communication data symbols can severely amplify noise background. To this end, the cyclic cross-correlation (CCC) is introduced in \cite{DFRC_SC_OFDM} to replace the steps i) and ii) of COS. For convenience, let us call the method developed in \cite{DFRC_SC_OFDM} as C-COS. Note that C-COS can be employed for sensing based on the orthogonal time-frequency space (OTFS) which is a potential waveform candidate for future mobile communications \cite{OTFS_Magazine_wei2020orthogonal}. 
While using C-COS requires OTFS to be cyclic prefixed for each symbol as in OFDM and \dso, OTFS with a reduced cyclic prefix (RCP), i.e., a single CP for the whole block of OTFS symbols, is the main trend in the OTFS literature. 

For \rcpotfs, COS and \ccos~cannot be directly applied. 
So far, \rcpotfs~sensing is mainly based on the maximum likelihood detection (MLD).
In \cite{OTFS_jcas2020twc}, an MLD problem for \rcpotfs~sensing is formulated in the delay-Doppler domain. In the case of a single target, 
solving an MLD can be done through a matching filter. This is performed in \cite{OTFS_yuan2021integratedSensingCOmmunicationOTFS} and \cite{OTFS_Raviteja2019TVT_embeddedPilotChannelEstimation}. 
In \cite{OTFS_keskin2021radarTimeDomainICIisi}, an MLD problem for \rcpotfs~sensing is formulated in the time domain which, as claimed therein, provides more insight compared with the MLD in the delay-Doppler domain. Regardless of the domains used for formulating MLD problems, solving them require an exhaustive search over the whole range-Doppler region. For each range-Doppler grid to be tested, a metric is calculated with high-dimensional matrix operations involved. 
In addition, solving the MLD problems requires a set of high-dimensional channel matrices which are pre-generated over the range-Doppler grids to be searched. 
Storing these matrices and accessing them in real time can be challenging in practice. 

In this paper, we develop a novel sensing framework that can be used for either waveforms with regular CPs, like OFDM and \dso, or waveforms with RCP, like \rcpotfs. 
The new sensing framework has a similar complexity to COS yet with enhanced sensing flexibility and performance. 
Our key innovations are illustrated below.

\begin{enumerate}[leftmargin=*]
	\item We propose a sensing framework that divides a block of signal evenly into multiple sub blocks.
	Unlike most existing schemes, such as COS, 
	we do not follow the underlying communication system and instead allow the number of samples in each sub-block to be different from that in a communication symbol. Moreover, we allow consecutive sub-blocks to overlap, which introduces a new flexibility 
	to optimize the sensing performance as well as to balance performance and implementation overhead.	
	
	\item We propose a virtual CP (VCP) that allows us to turn the echo signal in each sub-block into a sum of scaled and cyclically-shifted versions of a known signal. 
	This then allows us to remove the communication data symbols in the frequency domain and generate RDMs, as done in COS or \ccos. Moreover, the duration of the proposed VCP can be flexibly adjusted according to the maximum sensing distance. 
	Such flexibility is not owned by COS and its variants, as they strictly follow the underlying communication system. Further, it is worth noting that the flexibility of the proposed VCP also lies in that 
	it can be adjusted for better sensing performance.

\item We provide a comprehensive analysis of the interference-plus-noise (IN) terms of the RDMs obtained under the proposed sensing framework. We prove that the IN terms in both RDMs approximately conform to Gaussian distributions. The statistical properties of the distributions are also derived. Moreover, we derive the signal-to-IN ratio (SINR) of both RDMs and extrapolate the results to COS and \ccos. 
Further, we provide a holistic comparison between COS and the proposed sensing framework for both RDMs. The performance of the proposed sensing framework under the two RDMs is also analytically compared.

\end{enumerate}
Extensive simulations are provided to validate the proposed sensing framework and our analysis. 
In particular, we demonstrate that, as consistent with our analysis,
the proposed sensing framework always outperforms COS and \ccos~in low SNR regions where the upper limit of the region also matches the analytical result. 
We also confirm that the RDM obtained using CCC, as in \ccos~\cite{DFRC_SC_OFDM}, has a greater SINR than the RDM obtained based on the point-wise division, as in COS \cite{DFRC_dsss2011procIeee}, in low SNR regions; however, the former outperforms the later in high SNR regions. 
While the low-SNR observation is consistent with the results in \cite{DFRC_SC_OFDM}, the high-SNR result is unveiled for the first time. The critical value differentiating low and high SNR regions is also derived.

We remark that sensing based on standardized communication waveforms can also 
be performed using preambles. In \cite{DFRC_802p11ad2018TVT_Kumari,DFRC_OpportunisticRadar_80211ad_2018TSP}, different sensing methods are developed using the Golay complementary sequences (GCSs) in the preamble of IEEE 802.11ad communication signals. 
In \cite{DFRC_CommCentric_duggal2020doppler}, the Doppler resilience of
IEEE 802.11ad-based sensing is improved by incorporating  Prouhet-Thue-Morse sequences in the preamble. 
While these methods exploit the superb auto-correlation feature of GCSs for a high ranging performance, it can be non-trivial to adapt them for other communication standards. Moreover, as stated in \cite{DFRC_SC_OFDM}, using a whole block of communication signal for sensing can be more robust to interference and noise compared with only using preambles.
This paper is devoted to developing a flexible sensing framework that can be applied to most, if not all, communication systems. 
Thus, we use the payload signals as in COS and \ccos. The comparison between preamble- and payload-based sensing is out of the scope of this work. In fact, as they both have some unique advantages, their combination can be an interesting future work.

\section{Signal Model and Problem Statement}\label{sec: signal model}

In this section, the signal model of the considered ISAC scenario is first established based on the OTFS modulation. 
Then, COS is briefly reviewed. This further elicits several important issues that have not been effectively solved yet. 
Solutions to the issues will be developed in sequential sections.

\subsection{Signal Model}\label{subsec: signal model}

We consider that
a communication-only node is turned into an ISAC node by incorporating a sensing receiver. The communication signals are transmitted; meanwhile, the receiver collects target echo for sensing. As in most CC-ISAC work \cite{DFRC_dsss2011procIeee,DFRC_SC_OFDM,DFRC_802p11ad2018TVT_Kumari,DFRC_OpportunisticRadar_80211ad_2018TSP}, we ignore the self-interference, i.e., the signal leakage directly from the transmitter to the receiver, due to the implicit full-duplex operation. 
As for the communication waveform, we consider the OTFS modulation, not only because it is a 
a potential waveform candidate for future mobile communications \cite{OTFS_Magazine_wei2020orthogonal}, but also due to its capability in representing other common multi-carrier waveforms, e.g., OFDM and \dso.

Let $ d_{i}~(i=0,1,\cdots,I-1) $ denote the data symbols to be transmitted, where the data symbols are independently drawn from the same constellation, e.g., 64-QAM, and $ I $ denotes the total number of data symbols.
In OTFS modulation, the $ I $ numbers of data symbols are first placed in a two-dimensional delay-Doppler plane. Let the delay and Doppler dimensions be discretized into $ M $ and $ N $ grids, respectively. 
Denoting the time duration of $ M $ data symbols as $ T $, the sampling frequency along the Doppler dimension is then $ \frac{1}{T} $, which leads to a Doppler resolution of $ \frac{1}{NT} $. Since there are $ M $ grids along the delay dimension, the corresponding resolution is $ \frac{T}{M} $. 
In the OTFS modulation, the data symbols can be mapped from the delay-Doppler domain into the frequency-time domain via the following transform \cite{OTFS_windowDesignWeizhiQiang2021}, 
\begin{align} \label{eq: S[m,n]}
	& S[m,n] 
	= \frac{1}{\sqrt{MN}} \sum_{k=0}^{N-1}\sum_{l=0}^{M-1} d_{kM+l} e^{\mj2\pi(\frac{nk}{N}-\frac{mT}{M}l\df)}
\end{align}
where $ \mj $ denotes the imaginary unit.
Note that frequency and time are the dual domains of delay and Doppler, respectively. 

The frequency-time-domain signal $ S[m,n] $ is then transformed into the time domain by performing the IDFT w.r.t. $ m $ for each $ n $. This leads to
\begin{align}\label{eq: s[l,n]}
	s[l,n] = \sum_{m=0}^{M-1}S[m,n]\myDFT{M}{-ml},~\forall n,
\end{align}
where 
$ \myDFT{M}{-ml} $ denotes the DFT basis, as given by
\begin{align}\label{eq: DFT basis}
	\myDFT{a}{bc} = e^{-\mj\frac{2\pi bc}{a}}/\sqrt{a}.
\end{align}
{If the critical sampling is employed, which is typical in practice, then $ T\df $ in (\ref{eq: S[m,n]}) becomes one and moreover the IDFT performed in (\ref{eq: s[l,n]}) will cancel the $ l $-related transform in (\ref{eq: S[m,n]}).}
Treating $ l $ as the row index and $ n $ the column index, the signal $ s[l,n] $ will be transmitted column-by-column and in each column the entries $ l=0,1,\cdots,M-1 $ are transmitted sequentially. Before going through the digital-to-analog converter, some extra processing on $ s[l,n] $ would be necessary. 
To prevent the inter-symbol interference (ISI), cyclic prefix is generally used in multi-carrier transmissions. There are two types of CP in the OTFS literature. 

\textit{In the first type}, every $ M $ data symbols have a CP added \cite{OTFS_DasOFDMbasedOTFSIEEEaccess2021}, which is referred to as CP-OTFS hereafter. Let $ Q $ denote the number of samples in a CP.
Based on $ s[l,n] $ given in (\ref{eq: s[l,n]}), the signal to be transmitted can be given by
\begin{align}\label{eq: s tilde CP [i]}
	&\tilde{s}_{\mathrm{CP}}[i] = s\left[ \myModulo{\myModulo{i}{M+Q}-Q}{M},\myFloor{{i}/{{(M+Q)}}} \right],\nonumber\\
	&~~~~~~~~~~~~~~~~~~~~~~~~~~~~~~ i=0,1,\cdots,N(M+Q)-1,
\end{align}
where $ \myModulo{x}{y} $ takes $ x $ modulo $ y $ and $ \myFloor{x} $ rounds toward negative infinity. The indexes on the RHS of (\ref{eq: s tilde CP [i]}) indicate that every $ (M+Q) $ samples of $ \tilde{s}[i ] $ are obtained by copying the last $ Q $ samples from $ s[0,n],s[1,n],\cdots,s[M-1,n] $ at some $ n $ and pasting to the beginning. 

\textit{In the second-type CP}, the whole block of $ MN $ data symbols have a single CP added, which is known as the reduced CP-OTFS (RCP-OTFS) \cite{OTFS_Raviteja2019TVTpulseShapingRCP}. The signal to be transmitted in RCP-OTFS can be given by
\begin{align}\label{eq: s tilde RCP [i]}
	&\tilde{s}_{\mathrm{RCP}}[i] = s\left[ \myModulo{\tilde{i}}{M},\myFloor{{\tilde{i}}/{{M}}} \right],\nonumber\\
	\mathrm{s.t.}~&~\tilde{i} = \myModulo{i-Q}{MN},~i=0,1,\cdots,MN+Q-1,
\end{align}
where the indexes on the RHS indicate that the last $ Q $ samples from $ s[0,N-1],s[1,N-1],\cdots,s[M-1,N-1] $ are copied and pasted to the beginning of $ s[l,0] $. 
The CP-added signal will go through a digital-to-analog conversion (DAC) and other analog-domain processing, e.g., frequency up-conversion and power amplification etc., before being transmitted. A pulse-shaping filter is generally performed to reduce the out-of-band (OOB) emission. {For illustration convenience, we 
do not include the filter in the signal model. Nevertheless, practical pulse-shaping filters will be used in our simulations.}

As mentioned earlier, we consider a sensing receiver co-located with the communication transmitter. 
Therefore, it is reasonable to assume prefect synchronization and zero frequency offset for sensing. Consider $ P $ targets. The scattering coefficient, time delay and Doppler frequency of the $ p $-th target are denoted by $ \alpha_p $, $ \tau_p $ and $ \nu_p $, respectively. Let
$ \tilde{s}[i] $ be either $ \tilde{s}_{\mathrm{CP}}[i] $ or $ \tilde{s}_{\mathrm{RCP}}[i] $ (the set of $ i $ varies accordingly).
The target echo, as a sum of the scaled and delayed versions of $ \tilde{s}[i] $, can be modeled as
\begin{align}\label{eq: x[i]}
	& x[i] = \sum_{p=0}^{P-1}\tilde{\alpha}_p \tilde{s}\left[ i-\ml_p \right] e^{\mj 2\pi i \tilde{\mk}_p} + w[i], \nonumber\\
	&\mathrm{s.t.}~\tilde{\alpha}_p = \alpha_p e^{-\mj 2\pi\nu_p \tau_p }; ~ \ml_p= {\tau_p}/{T_{\mathrm{s}}};~ \tilde{\mk}_p = \nu_p\ts,
\end{align}
where $ \ts $ is the sampling interval, and $ w[i]\sim\cn{\sigma_w^2} $ is the additive noise conforming to a circularly-symmetric complex centered Gaussian distribution.

\begin{remark}\label{rmk: otfs represents dft-s-ofdm and ofdm}	
	Some features of the above signal model are remarked here. \textit{First}, the signal $ \scpi $ can represent \dso~and OFDM with slight changes made on $ S[m,n] $. In particular, \dso~can be obtained when the Fourier transform w.r.t. $ k $ is suppressed in (\ref{eq: S[m,n]}), while OFDM is obtained when both Fourier transforms in (\ref{eq: S[m,n]}) are skipped. 
	\textit{Second}, $ s[l,n] $ obtained in (\ref{eq: s[l,n]}) approximately conforms to a complex centered Gaussian distributions, as denoted by $ s[l,n]\sim \mathcal{CN}(0,\sigma_d^2) $, where $ \sigma_d^2 $ is the power of $ d_i $; see (\ref{eq: S[m,n]}).  
	The above result can be attained using (\ref{eq: S[m,n]}) and (\ref{eq: s[l,n]}) in combination with another two facts: the complex envelope of an uncoded OFDM system converges in distribution to a complex Gaussian random process \cite{Gaussian_OFDMenvelope}; the unitary DFT does not change the statistical properties and the whiteness of a Gaussian process \cite{book_oppenheim1999discrete}. 
	\textit{Third}, we can use the above two facts to validate that $ s[l,n]\sim \mathcal{CN}(0,\sigma_d^2) $ also holds for \dso~and OFDM.	
\end{remark}

\subsection{Classical OFDM Sensing (COS)}\label{subsec: reviewing classical ofdm sensing}

COS was developed about a decade ago and has been widely used in the sensing literature; see \cite{OFDM_autonomousDriv2019microwaveMag,BinYang_ofdmSPmagazine} and their references. However, there are still some issues that have not been effectively solved yet. 
Below, we briefly review COS and highlight the issues.
As COS is originally developed for CP-OFDM, we assume that $ \scpi $ is transmitted and use $ x[i] $ to describe the method.

In COS, the $ (M+Q)N $ numbers of echo samples $ x[i] $ are divided into $ N $ consecutive symbols, each having $ (M+Q) $ samples. 
Removing the first $ Q $ samples in each symbol and taking the $ M $-point unitary DFT of the remaining samples in the symbol, we obtain a common echo signal model 
\begin{align}\label{eq: Xn[m]}
	X_n[m] \approx \sum_{p=0}^{P-1}\tilde{\alpha}_p S[m,n]  e^{-\mj \frac{2\pi m \ml_p }{\bar{M}}}  e^{\mj 2\pi n(M+Q)\tilde{\mk}_p}  +  {W}_n[m], \nonumber\\[-5mm]
\end{align}
where $ S[m,n] $ is the signal given in (\ref{eq: S[m,n]}) but with the two Fourier transforms suppressed, as illustrated in Remark \ref{rmk: otfs represents dft-s-ofdm and ofdm}, and $ W_n[m] $ denotes the DFT of the background noise. Note that, as often done in radar signal processing \cite{book_richards2010principlesModernRadar}, the intra-symbol Doppler effect is suppressed in (\ref{eq: Xn[m]}). 

Dividing $ X_n[m] $ by $ S[m,n] $ in a point-wise manner, we can remove the communication data symbols. Then, a two-dimensional Fourier transform can be performed over $ m $ and $ n $, leading to the following range-Doppler map (RDM),
\begin{align} \label{eq: U r k[l]}
	& U_k^{\mathrm{r}}[l] = \sum_{n=0}^{N-1}\sum_{m=0}^{M-1} X_n[m]\Big/S[m,n] \myDFT{M}{-ml}\myDFT{N}{nk},\\
	& = \sum_{p=0}^{P-1}{\tilde{\alpha}_p} \mySinc{{M}}{l-\ml_p} \mySinc{{N}}{(M+Q)N\tilde{\mk}_p-k} + {W}_k^{\mathrm{r}}[l],\nonumber
\end{align}
where the superscript $ \{\cdot\}^{\mathrm{r}} $ stands for `ratio' to differentiate with another way of removing $ S[m,n] $, as to be illustrated in Section \ref{subsubsec: signal cancellation}; $ \myDFT{a}{bc} $ is the unitary DFT basis and defined in (\ref{eq: DFT basis}); $ {W}_k[l] $ is the two-dimensional Fourier transform of $ W_n^{\mathrm{r}}[m]\Big/S[m,n] $; and $ \mySinc{x}{y} $ is introduced to denote the DFT results of the two exponential signals in (\ref{eq: Xn[m]}). The general form of $ \mySinc{x}{y} $ is given by
\begin{align} \label{eq: Sinc definition}
	\mySinc{x}{y} 
	= \frac{1}{\sqrt{x}}\frac{\sin\left( \frac{x}{2} \frac{2\pi y}{x} \right)}{\sin\left( \frac{1}{2} \frac{2\pi y}{x} \right) }e^{\mj \frac{x-1}{2} \frac{2\pi y}{x} }.
\end{align}
The function $ \mySinc{x}{y}  $ is localized around $ y=0 $ and hence 
$ |Y_k[l]| $ can present $ P $ dominant peaks in the range-Doppler domain, if $ \tilde{\alpha}_p~{\forall p} $ is sufficiently large. 
Thus, a threshold detector based on, e.g., likelihood ratio test (LRT), can be developed for target detection, from which coarse estimations of target parameters can also be attained.

\subsection{Motivation and Problem Statement}\label{subsec: problem statement}

COS has been widely applied given its low complexity. However, COS and many of its variants can have limited sensing performance, as they follow the underlying communication systems. 
Some intriguing issues are illustrated below.

\subsubsection{CP-limited sensing distance}\label{subsubsec: CP limitation}

CP plays some non-trivial role in COS. {Specifically, 
	CP makes each received symbol consists of cyclically shifted version of the transmitted symbol. This then enables us to attain the convenient echo model given in (\ref{eq: Xn[m]}) and further facilitates the removal of $ S[m,n] $ to generate the RDM given in (\ref{eq: U r k[l]}).} 
However, CP also puts a constraint on sensing. Namely, 
the round-trip delay of the maximum sensing distance should be smaller than the time duration of the CP. 
Such limitation stands even when we have a sufficient link budget for sensing a longer distance. 
Moreover, for the communication waveform with a reduced CP, as modeled in (\ref{eq: s tilde RCP [i]}), 
COS is not directly applicable. 
	
	\subsubsection{Communication-limited velocity measurement}\label{subsubsec: M and N selection}
	While the sensing distance is limited by CP, the velocity measurement performance can be constrained by the values of $ M $ and $ N $. 
	Substituting $ \tilde{k}_p=\nu_p T_{\mathrm{s}} $ into (\ref{eq: Xn[m]}), we see that the Doppler frequency $ \nu_p $
	becomes the frequency of the exponential signal of $ n $ and $ (M+Q)T_{\mathrm{s}} $ is the sampling interval. Thus, the maximum (unambiguous) measurable value of the Doppler frequency, as denoted by $ \nu_{\mathrm{max}} $, and its resolution, as denoted by $ \dnu $, can be given by 
	 \begin{align} 
	 	\nu_{\mathrm{max}} = {1}\Big/\big(2(M+Q)T_{\mathrm{s}}\big);~
	 	\dnu = {1}\Big/\big(N(M+Q)T_{\mathrm{s}}\big). \nonumber
	 \end{align}
	While a small $ M $ can give us a large unambiguous region for Doppler measurement, a large $ N $ is then necessary to keep a small $ \dnu $. 
	However, assigning the values of $ M $ and $ N $ in a sensing-favorable way may degrade the performance of the underlying communication system, e.g., 5G \cite{book_ahmadi2019_5G}, that generally has stringent requirements on the two parameters.

\subsubsection{COS adapted for \dso}\label{subsubsec: signal cancellation}

	As shown in (\ref{eq: U r k[l]}), communication data symbols are removed via point-wise divisions in COS. For CP-OFDM, this is okay, as $ S[m,n] $, directly drawn from a constellation, does not take zero in general. However, for \dso~and OTFS, 
	$ S[m,n] $ conforms to a complex centered Gaussian distribution, as illustrated in Remark \ref{rmk: otfs represents dft-s-ofdm and ofdm}. This means a certain portion of $ S[m,n] $ is centered around the origin and the direct division can lead to severe noise amplification. 
	To address the issue, a time-domain cyclic cross-correlation (CCC) is proposed in \cite{DFRC_SC_OFDM} to replace the frequency-domain division. The RDM under CCC can be written based on (\ref{eq: U r k[l]}), leading to
	\begin{align} \label{eq: U c k[l]}
		& U_k^{\mathrm{c}}[l] = \sum_{n=0}^{N-1}\sum_{m=0}^{M-1} X_n[m]S^*[m,n] \myDFT{M}{-ml}\myDFT{N}{nk},
	\end{align}
	where a closed-form result, as in the second line of (\ref{eq: U r k[l]}), is not available, due to the randomness of $ S^*[m,n] $. Note that $ S[m,n]  $ here is not the same as in (\ref{eq: U r k[l]}). As said in Remark \ref{rmk: otfs represents dft-s-ofdm and ofdm}, for \dso, $ S[m,n] $ can be obtained by suppressing the $ k $-related Fourier transform in (\ref{eq: S[m,n]}). 
	Now that we have two ways of generating RDMs, a question 
	follows naturally: which one gives the better sensing performance? 

{It is worth noting that the issues highlighted in Sections \ref{subsubsec: CP limitation} and \ref{subsubsec: M and N selection} have barely been treated yet in the literature. As seen in \cite{OFDM_autonomousDriv2019microwaveMag,BinYang_ofdmSPmagazine,DFRC_SC_OFDM,DFRC_dsss2011procIeee}, COS and many of its variants often 
	follow the underlying communication system w.r.t. $ M $ and $ N $ and
	assume by default that the maximum round-trip delay of sensing targets is no greater than the CP duration. Moreover, the question asked in Section \ref{subsubsec: signal cancellation} has not been systematically investigated, although it was shown through simulations in \cite{DFRC_SC_OFDM} that the CCC-based RDM has better sensing performance than the ratio-based RDM in low SNR regions. This, however, is not always the case, as will be unveiled in Section \ref{sec: analyze and compare RDMs}. 
	To address the issues highlighted in Sections \ref{subsubsec: CP limitation} and \ref{subsubsec: M and N selection}, we develop a novel flexible sensing framework in Section \ref{sec: sensing framework}. Performance analysis for the proposed sensing framework will be conducted in Section \ref{sec: analyze and compare RDMs}, which also answers the question asked in Section \ref{subsubsec: signal cancellation}.}

\section{A Low-Complexity Sensing Framework} \label{sec: sensing framework}

The proposed sensing framework starts with segmenting the whole block of sensing-received samples into multiple consecutive sub-blocks, then removes the communication data symbols in the frequency domain, and finally generates an RDM. 
Substantially differentiating the proposed sensing framework from COS is the way a block of samples is segmented, which is 
detailed next.

\begin{figure}[!t]
	\centering 
	\includegraphics[width=88mm]{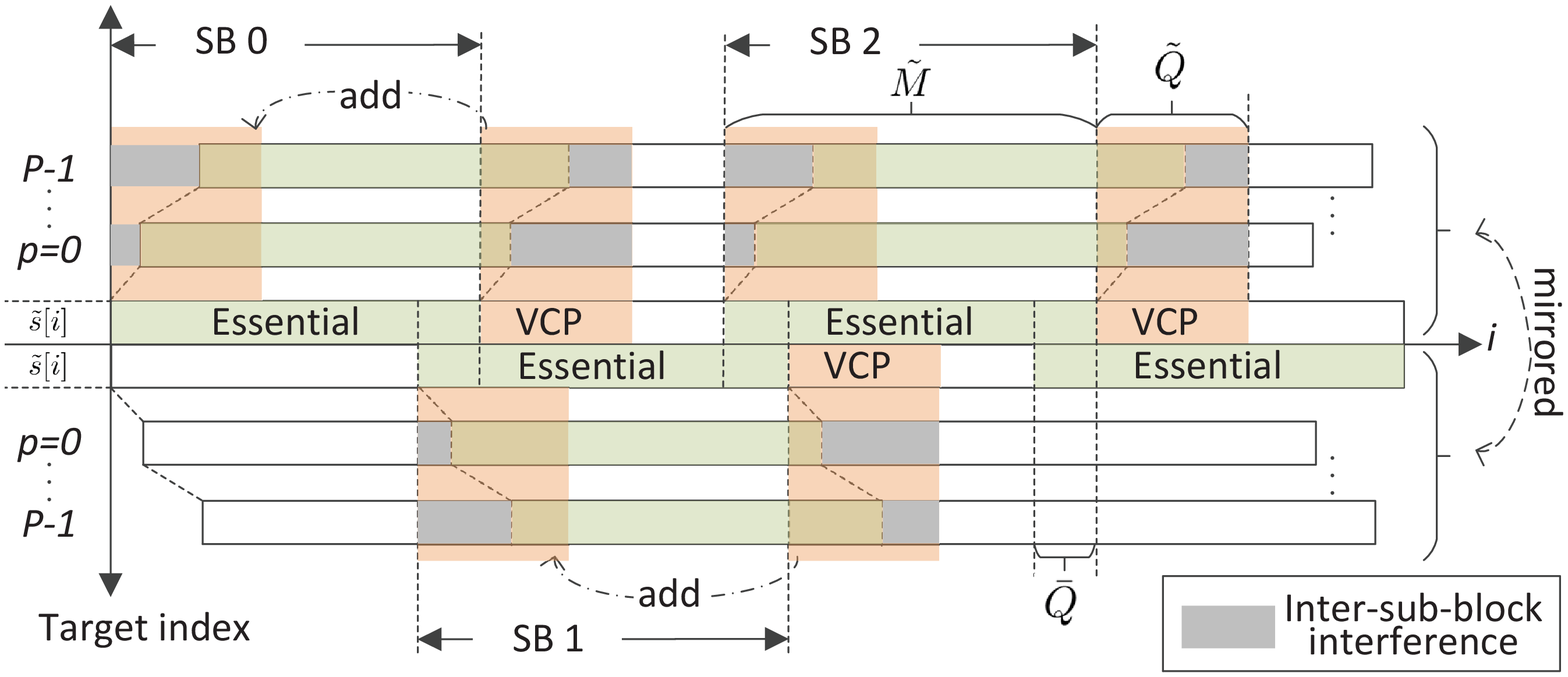}
	\caption{Illustrating the proposed block segmentation and virtual cyclic prefix (VCP), where $ p $ is the target index. 
		The received signal $ x[i] $ is the sum of the scaled versions of the delayed signals under $ p=0,1,\cdots,P-1 $. 
	The proposed block segmentation divides the whole block into multiple sub-blocks (SBs), each having $ \tilde{M} $ samples. Adjacent SBs overlap by $ \bar{Q} $ samples. The proposed VCP adds the $ \tilde{Q} $ samples right after each SB onto the beginning of the SB, turning a SB into the sum of cyclic shifted versions of the essential part of the SB. The cost of adding VCP, however, is inter-SB interference, which is analyzed in Section \ref{sec: analyze and compare RDMs}.}
	\label{fig: rearrange and VCP}
	
	\vspace{-5mm}
\end{figure}

As illustrated in Fig. \ref{fig: rearrange and VCP}, we propose to segment $ x[i] $ into $ \tilde{N} $ sub-blocks, each having $ \tilde{M} $ samples, where $ \tilde{N}=N $ is not required in our design.
Moreover, we allow the consecutive sub-blocks to overlap by $ \bar{Q} $ samples, where $ \bar{Q} $ is either zero or a positive integer. 
As also shown in Fig. \ref{fig: rearrange and VCP}, we segment the communication-transmitted signal $ s[i] $ in the same way as described above and call each segment the essential signal of the sub-block. 
Due to the propagation delay of a target, part of the essential signal is not within the received sub-block but right after it. To preserve the essential signal in each sub-block, we propose to add the $ \tilde{Q} $ samples right after a sub-block onto the first $ \tilde{Q} $ samples within the sub-block, creating a virtual CP (VCP). As seen from Fig. \ref{fig: rearrange and VCP}, adding VCP can make each received sub-block comprised of cyclically shifted versions of its essential signal part, as long as $ \tilde{Q} $, \textit{not $ Q $ any more}, is greater than the maximum target delay. 
Since the value of $ \tilde{Q} $ is not limited to the original CP length $ Q $, we can design the maximum sensing distance flexibly subject to a sufficient link budget. Next, the above description is further elaborated on using the signal model provided in Section \ref{sec: signal model}.

Let $ s_n[l] $ denote the essential signal of the $ n $-th sub-block. Based on the above illustration, we can write $ s_n[l] $ as
\begin{align} \label{eq: s n[l]}
	& {s}_n[l] = \tilde{s}[n(\tilde{M}-\bar{Q})+l],~l=0,1,\cdots,\tilde{M}-1,\nonumber\\
	&~~~~~~~~~~ n=0,1,\cdots,\tilde{N}-1,~\tilde{N} = \myFloor{ \frac{(I-\tilde{Q}-\bar{Q})}{(\tilde{M}-\bar{Q})} },
\end{align}
where 
$ \tilde{s}[\cdot] $ on the RHS can be either $ \scpi $ in (\ref{eq: s tilde CP [i]}) or $ \srcpi $ in (\ref{eq: s tilde RCP [i]}),
$ \tilde{N} $ is the total number of sub-blocks\footnote{
	Take the three sub-blocks in 
	Fig. \ref{fig: rearrange and VCP} for an illustration.
	By excluding the last $ \bar{Q}$
	samples of sub-block two and its $ \tilde{Q} $-sample VCP, we see that each of the first three sub-blocks has $ (\tilde{M}-\bar{Q}) $ unique samples. This can be generalized into the expression of $ \tilde{N} $ given in (\ref{eq: s n[l]}).  
} and $ \myFloor{\cdot} $ rounds towards negative infinity.
With reference to Fig. \ref{fig: rearrange and VCP}, after adding VCP, the received signal in sub-block $ n  $ becomes
\begin{align} %
	& {x}_n[l]  \approx \sum_{p=0}^{P-1}\tilde{\alpha}_p  {s}_n\left[ \myModulo{l- \ml_p }{\tilde{M}} \right] e^{\mj 2\pi n(\tilde{M}-\bar{Q})\tilde{\mk}_p} + {w}_n[l] +\nonumber\\
	&~~~~~~~~~~~  ~~~~~~ z_n^{(p)}[l]g_{\tilde{Q}}[l], ~l=0,1,\cdots,\tilde{M}-1,
	\label{eq: x n[l]}
\end{align}
where $ \tilde{\alpha}_p $, $ \ml_p $ and $ \tilde{\mk}_p $ are given in (\ref{eq: x[i]}). {Similar to (\ref{eq: Xn[m]}), the approximation here is also due to the suppression of the intra-sub-block Doppler impact.} We emphasize that, due to the $ \bar{Q} $-sample overlapping of consecutive sub-blocks, the Doppler phase is $ 2\pi(\tilde{M}-\bar{Q})\tilde{\mk}_p $ \textit{not} $ 2\pi \tilde{M}\tilde{\mk}_p $. 
In (\ref{eq: x n[l]}), $ z_n^{(p)}[l]g_{\tilde{Q}}[l] $ denotes the interference term and $ g_{\tilde{Q}}[l] $ is a rectangular window function which takes one at $ l=0,1,\cdots,\tilde{Q}-1 $, and zero elsewhere. 
Moreover, the noise term $ w_n[l] $ in (\ref{eq: x n[l]}) is obtained by first segmenting $ w[i] $ given in (\ref{eq: x[i]}) as done in (\ref{eq: s n[l]}) and 
then adding VCP. 
Since the addition of two i.i.d. Gaussian variables is still Gaussian with the variance doubled, we have
\begin{align} \label{eq: w n[l]}
	{w}_n[l]\sim \left\{  
		\begin{array}{ll}
			\mathcal{CN}(0,2\sigma_w^2)&\mathrm{~for~}l=0,1,\cdots,\tilde{Q}-1\\
			\mathcal{CN}(0,\sigma_w^2)&\mathrm{~for~}l=\tilde{Q},\cdots,\tilde{M}-1
		\end{array}	
	\right..
\end{align}

As in OFDM, the cyclic shift of the essential signal preserves the sub-carrier orthogonality. Therefore, taking the $ \tilde{M} $-point DFT of $ x_n[l] $ w.r.t. $ l $ leads to
\begin{align}\label{eq: X n[m] new sensing}
	&{X}_n[m] = \sum_{p=0}^{P-1}\tilde{\alpha}_p {S}_n[m]  e^{-\mj \frac{2\pi m \ml_p }{\tilde{M}}}  e^{\mj \frac{2\pi n{\mk}_p}{\tilde{N}}}  +  {W}_n[m] + Z_n[m], \nonumber\\
	&
	\mathrm{s.t.}~{\mk}_p = \tilde{N}(\tilde{M}-\bar{Q})\tilde{\mk}_p,\nonumber\\
	&~~~~ F_n[m]=\sum_{l=0}^{\tilde{M}-1}f_n[l]\myDFT{\tilde{M}}{lm},~(F,f)\in \left\{{(S,s),(W,w)} \right\},\nonumber\\
	&~~~~ Z_n[m] = \sum_{p=0}^{P-1}\tilde{\alpha}_p \sum_{l=0}^{\tilde{M}-1} z_n^{(p)}[l]g_{\tilde{Q}}[l]  \myDFT{\tilde{M}}{lm},
\end{align}
where $ {S}_n[m] $, $ {W}_n[m] $ and $ Z_n[m] $ are the DFTs of the respective terms in (\ref{eq: x n[l]}). We notice again that the unitary DFT basis, as defined in (\ref{eq: DFT basis}), is used.
Since $ s_n[l] $ is known, $ S_n[m] $ can be readily calculated. 
Corresponding to (\ref{eq: U r k[l]}), we can divide both sides of $ {X}_n[m] $ by $ S_n[m] $ and take the two-dimensional DFT w.r.t. $ n $ and $ m $, attaining the following ratio-based RDM:
\begin{align} \label{eq: V k r l our rdm}
	& V_k^{\mathrm{r}}[l] = \sum_{p=0}^{P-1}{\tilde{\alpha}_p} \mySinc{\bar{M}}{l-\ml_p} \mySinc{\tilde{N}}{\mk_p-k} + {W}_k^{\mathrm{r}}[l] + Z_k^{\mathrm{r}}[l],\nonumber\\
	&\mathrm{s.t.}~\mX_k^{\mathrm{r}}[l] = \sum_{n=0}^{N-1}\sum_{m=0}^{M-1} \frac{\mX_n[m]}{S_n[m]} \myDFT{M}{-ml}\myDFT{N}{nk},~\mX\in \left\{W,Z \right\}
\end{align}
where $ \mySinc{x}{y} $ is defined in (\ref{eq: Sinc definition}).
Corresponding to (\ref{eq: U c k[l]}), we can multiply both sides of $ X_n[m] $
with the conjugate of $ S_n[m] $ and take the same DFT w.r.t. $ n $ and $ m $ as above, obtaining the CCC-based RDM:
\begin{align}\label{eq: V k c l our rdm}
	& V_k^{\mathrm{c}}[l] = \sum_{p=0}^{P-1}\tilde{\alpha}_p  S_k^{\mathrm{c}}[l] + {W}_k^{\mathrm{c}}[l] + Z_k^{\mathrm{c}}[l],\\
	&\mathrm{s.t.}~S_k^{\mathrm{c}}[l] = \sum_{n=0}^{\tilde{N}-1}\sum_{m=0}^{\tilde{M}-1}
	|{S}_n[m]|^2  e^{-\mj \frac{2\pi m \ml_p }{\tilde{M}}}  e^{\mj \frac{2\pi n{\mk}_p}{\tilde{N}}}\myDFT{\tilde{M}}{-ml}\myDFT{\tilde{N}}{nk};\nonumber\\
	& \mX_k^{\mathrm{r}}[l] = \sum_{n=0}^{\tilde{N}-1}\sum_{m=0}^{\tilde{M}-1} {\mX_n[m]}{S_n^*[m]} \myDFT{\tilde{M}}{-ml}\myDFT{\tilde{N}}{nk},~\mX\in \left\{W,Z \right\}. \nonumber
\end{align}

{Based on the RDMs obtained in (\ref{eq: V k r l our rdm}) and (\ref{eq: V k c l our rdm}), target detection and parameter estimation can be performed for various sensing applications. Note that developing methods for target detection and parameter estimation will be out of the scope of this study, as we focus on designing the framework and investigating the unsolved issues highlighted in Section \ref{subsec: problem statement}. 
Nevertheless, as will be proved in the next section, Propositions \ref{pp: interference and noise of ratio-RDM} and \ref{pp: signal components in RDM ccc} in specific, the IN signals in both RDMs, i.e., $ {W}_k^{\mathrm{r}}[l] + Z_k^{\mathrm{r}}[l] $ and $ {W}_k^{\mathrm{c}}[l] + Z_k^{\mathrm{c}}[l] $, over range-Doppler grids, i.e., $ k $ and $ l $, approximately conform to i.i.d. Gaussian distributions. 
This enables many existing target detectors and parameter estimators to be directly applicable under the proposed sensing framework. 
To validate the new design and analysis, the cell-averaging constant false-alarm rate detector (CA-CFAR) \cite[Chapter 16]{book_richards2010principlesModernRadar} will be performed in our simulations.
}

\begin{table}[!t]%
	\caption{\small Proposed Sensing Framework}
	\vspace{-3mm}	
	\begin{center}\footnotesize
		\begin{tabular}{p{8.5cm}}
			\hline
			\vspace{0.5mm}
			\textit{Input}: $ \tilde{M} $, $ \tilde{Q} $, $ \bar{Q} $, $ S_n[m] $ and $ \tilde{N} $ given in (\ref{eq: s n[l]}), and $ x[i] $ given in (\ref{eq: x[i]}).
			
			\begin{enumerate}[leftmargin=*]\renewcommand{\labelenumi}{{\arabic{enumi})}}
				
				\item Segment $ x[i] $ into $ \tilde{N} $ sub-blocks (SBs): the $ n $-th SB starts from the $ n(\tilde{M}-\bar{Q})~(n=0,1,\cdots,\tilde{N}-1) $ and has $ \tilde{M} $ samples;  
				
				\item Add the $ \tilde{Q} $ samples after each SB onto the first $ \tilde{Q} $ within the SB;
				
				\item Take the $ \tilde{M} $-point DFT of each SB, attaining $ X_n[m] $ given in (\ref{eq: X n[m] new sensing});
				
				\item If the ratio-based RDM is preferred, divide $ X_n[m] $ by $ S_n[m] $ pointwise and take a two-dimensional DFT w.r.t. $ n $ and $ m $, leading to (\ref{eq: V k r l our rdm});
				
				\item  If the CCC-based RDM is chosen, multiply $ X_n[m] $ with 
				$ S_n^*[m] $ pointwise and take a two-dimensional DFT, yielding (\ref{eq: V k c l our rdm});
				
				\item Provided $ \mP_{\mathrm{F}} $, $ N_{\mathrm{g}}^k $, $ N_{\mathrm{g}}^l $, $ N_{\mathrm{r}}^k $ and $ N_{\mathrm{r}}^l $, enumerate each range-Doppler grid by performing the following steps, 
				
				\begin{enumerate}
					\item Estimate the power of the local IN background according to (\ref{eq: sigma hat k* l*}); 
					
					\item Calculate the detecting threshold $ \mT $ based on (\ref{eq: T threshold});
					
					\item If a power of the grid under test is greater than $ \mT $, a target exists; otherwise, no target.
					If a target exists, the coarse estimates of its parameters can be obtained; see (\ref{eq: tau hat p, nu hat p}).
				\end{enumerate}
				
								\vspace{-3mm}
			\end{enumerate}\\
			\hline
		\end{tabular}
				\vspace{-5mm}
	\end{center}
	\label{tab: proposed sensing framework}
\end{table}	

Next, we summarize the proposed sensing framework in Table \ref{tab: proposed sensing framework}, where CA-CFAR is also briefly described.
From the input of Table \ref{tab: proposed sensing framework}, we see some extra parameters, e.g., $ \tilde{M} $, $ \tilde{Q} $ and $ \bar{Q} $ that are not owned by COS. 
These parameters endow the proposed sensing framework with better flexibility and adaptability compared with COS. Their design criteria will be illustrated in Section \ref{sec: analyze and compare RDMs}. In Table \ref{tab: proposed sensing framework},
Steps 1) to 2) perform the proposed block segmentation and VCP. 
Step 3) transforms the time-domain signal into the frequency domain. Steps 4) and 5) show two different ways of removing communication data symbols and accordingly generate RDMs. 
Step 6) and its sub-steps implement the CA-CFAR. 

In Step 6), $ \mP_{\mathrm{F}} $
is the expected false-alarm rate; $ N_{\mathrm{g}}^k $ and $ N_{\mathrm{g}}^l $ denote the number of gap samples on each side of the grid under test (GUT) along the $ k $- and $ l $-dimensions; likewise, $ N_{\mathrm{r}}^k $ and $ N_{\mathrm{r}}^l $ denote the number of reference samples. The gap samples will be excluded while the reference samples will be used, when 
estimating the power of local IN background.  
Given a Gaussian IN background, the maximum likelihood estimate of the power is the mean of the signal power of the selected reference grids, i.e., 
\begin{align} \label{eq: sigma hat k* l*}
	& \hat{\sigma}_{k^*,l^*}^2 = \frac{1}{|\Omega^{\mathrm{r}}_{k^*,l^*}|}\sum_{(k,l)\in \Omega^{\mathrm{r}}_{k^*,l^*}} |V_{k}^{\mX}[l]|^2 ,~\mX\in \{\mathrm{r},\mathrm{c}\}, \nonumber\\
	& \Omega^{\mathrm{r}}_{k^*,l^*}=\left\{ (k,l)\Big| \substack{k=k^*-N_{\mathrm{r}}^k-N_{\mathrm{g}}^k,\cdots, k^*+N_{\mathrm{r}}^k+N_{\mathrm{g}}^k;
		\\
		l=l^*-N_{\mathrm{r}}^l-N_{\mathrm{g}}^l,\cdots, l^*+N_{\mathrm{r}}^l+N_{\mathrm{g}}^l } \right\} \Big\backslash \nonumber\\
	&~~~~~~~~~~~~~~~~~~~~~~~~~~~~ \left\{ (k,l)\Big| \substack{k=k^*-N_{\mathrm{g}}^k,\cdots, k^*+N_{\mathrm{g}}^k;
		\\
		l=l^*-N_{\mathrm{g}}^l,\cdots, l^*+N_{\mathrm{g}}^l } \right\},
\end{align}
where $ (k^*,l^*) $ denotes the index of GUT, $ \Omega^{\mathrm{r}}_{k^*,l^*} $ denotes the index set of reference grids, $ \{\}\backslash\{\} $ gives the set difference, and $ |\Omega| $ denotes the number of entries in the set $ \Omega $. Using $ \hat{\sigma}_{k^*,l^*}^2 $, we can set the CA-CFAR threshold as \cite[(16.23)]{book_richards2010principlesModernRadar},
\begin{align}\label{eq: T threshold}
	\mT=\beta \hat{\sigma}_{k^*,l^*}^2, ~\beta = |\Omega^{\mathrm{r}}_{k^*,l^*}|\left( \mP_{\mathrm{F}}^{-1/|\Omega^{\mathrm{r}}_{k^*,l^*}|} - 1 \right).
\end{align}
If $ |V_{k^*}^{\mX}[l^*]|^2\ge \mT $, we report the presence of a target at $ (k^*,l^*) $. The coarse estimates of the delay and Doppler frequency of the target, say the $ p $-th, can be obtained as
\begin{align} \label{eq: tau hat p, nu hat p}
	\hat{\tau}_p = l^*\ts;~\hat{\nu}_p = k^*\Big/\left( (\tilde{M}-\bar{Q})\tilde{N}\ts \right),
\end{align}
where the relationship among relevant variables, as given in (\ref{eq: x[i]}) and (\ref{eq: X n[m] new sensing}), is used for the above result. {For applications requiring high-accuracy estimations of target
 location and velocity, various methods for parameter refinement are available in the literature, such as 
the conventional multiple signal clarification (MUSIC) \cite{book_van2004optimum} and a much newer 
DFT-interpolation-based estimator \cite{Kai_padeFreqEst2021TVT} etc. 
Details are suppressed here for brevity.
}

{Before ending the section, the last note is given on the computational complexity of the proposed sensing framework. From Table \ref{tab: proposed sensing framework}, we see that the computations performed in Steps 3)-6) dominate the overall complexity. Step 3) has the complexity of $ \mathcal{O}\{\tilde{N}\tilde{M}\log\tilde{M}\} $, where $ \mathcal{O}\{\tilde{M}\log\tilde{M}\} $ is complexity of the $ \tilde{M} $-point DFT (under the fast implementation \cite{Kai_ofdmSensingSPM}). The complexity of generating an RDM, performing either Step 4) or Step 5), is dominated by the two-dimensional DFT and can be given by $ \mathcal{O}\{\tilde{N}\tilde{M}\log\tilde{M}+\tilde{M}\tilde{N}\log\tilde{N}\} $. Step 6) essentially processes the RDM by a two-dimensional filter, and hence can be performed through the same two-dimensional Fourier transform as in Steps 4) and 5)
\cite{CFAR_FFTimplement_kronauge2013fast}. 
Consequently, we can say that the overall computational complexity of the proposed sensing framework is $ \mathcal{O}\{\tilde{N}\tilde{M}\log\tilde{M}+\tilde{M}\tilde{N}\log\tilde{N}\} $. 
}

\section{Performance Analysis} \label{sec: analyze and compare RDMs}

In this section, we analyze the interference and noise background in the two RDMs obtained in (\ref{eq: V k r l our rdm}) and (\ref{eq: V k c l our rdm}). Then we derive, analyze and compare their SINRs, through which the question in Section \ref{subsubsec: signal cancellation} will be answered. Moreover, insights into the parameter design for the proposed sensing framework will also be drawn.

\subsection{Preliminary Results} \label{subsec: preliminary results}

From (\ref{eq: V k r l our rdm}) and (\ref{eq: V k c l our rdm}), we see that both RDMs are obtained based on $ X_n[m] $ given in (\ref{eq: X n[m] new sensing}). Thus, we analyze first its three signal components, i.e., $ S_n[m],~Z_n[m] $ and $ W_n[m] $. 
Their useful features are provided here. In particular, their distributions are provided in Lemma \ref{lm: Sn[m] Wn[m] Zn[m] distributions} with the proof given in Appendix \ref{app: proof of lemma on S W Z distriutions}. 
The independence of $ S_n[m],~Z_n[m] $ and $ W_n[m] $ over $ n $ is given in Lemma \ref{lm: Sn[m] Wn[m] Zn[m] dependence}; see Appendix \ref{app: proof of lemma on Sn[m] Wn[m] Zn[m] dependence} for the proof. In addition, the independence of the signals over $ m $ is illustrated in Lemma \ref{lm: Zn[m] dependence over m}; see Appendix \ref{app: proof of lemma on dependence Zn[m] over m} for the proof.

\mySpaceTwoMM

\begin{lemma}\label{lm: Sn[m] Wn[m] Zn[m] distributions}
	{\it
	The useful signal and the noise in (\ref{eq: X n[m] new sensing}) satisfy
	\begin{equation}
		\begin{gathered}\label{eq: S n[m] CN, W n[m] CN}
			S_n[m]\sim\mathcal{CN}(0,\sigma_d^2);\\
			W_n[m]\sim\myCN{0,\sigma_{W}^2},~\sigma_W^2 = \left( 1 + {\tilde{Q}}/{\tilde{M}} \right)\sigma_w^2,
		\end{gathered}
	\end{equation}
where $ \sigma_d^2 $ is the power of communication data symbols, i.e., $ d_i $ given in (\ref{eq: S[m,n]}), and $ \sigma_w^2 $ is power of the receiver noise, i.e., $ w[i] $ given in (\ref{eq: x[i]}).
Moreover, provided that $ \alpha_0,\alpha_1,\cdots,\alpha_{P-1} $ are uncorrelated,
the interference term in (\ref{eq: X n[m] new sensing}) conforms to
	\begin{align} \label{eq: Z n[m] CN} Z_n[m]\sim\myCN{0,\sigma_{Z}^2},~\sigma_Z^2=\frac{{\tilde{Q}\sigma_d^2\sigma_P^2}}{\tilde{M}},~\sigma_P^2=\sum_{p=0}^{P-1}\sigma_p^2,
	\end{align}
where 
$ \sigma_p^2 $ is the power of the $ p $-th scattering coefficient, i.e., $ \alpha_p $ given in (\ref{eq: x[i]}). 
}
\end{lemma}

\mySpaceTwoMM

\begin{lemma} \label{lm: Sn[m] Wn[m] Zn[m] dependence}
	{\it 
		Given $ \tilde{M}>(\tilde{Q}+\bar{Q}) $ and at any $ m $,
	$ Z_n[m] $ is i.i.d. over $ n $, whereas $ S_n[m] $ and $ W_n[m] $ are each independent over {either the set of odd $ n $'s 
	or that of even $ n $'s}. In addition, we have, at any $ m $, 
	\begin{gather}
		\myCC{S_n[m],S_{n+1}[m]} = \bar{Q}/\tilde{M};\nonumber \\
		\myCC{W_n[m],W_{n+1}[m]} = (\tilde{Q}+\bar{Q})/\tilde{M}.\label{eq: C(SS) C(SZ) C(WW) C(ZZ)}
	\end{gather}	
where $ n=0,1,\cdots,\tilde{N}-2 $ and $ \myCC{x,y} = \frac{|\myExp{xy^*}|}{\sqrt{\myExp{|x|^2}\myExp{|y|^2}}} $ is the absolute correlation coefficient between $ x $ and $ y $.
}
\end{lemma}

\mySpaceTwoMM

\begin{lemma} \label{lm: Zn[m] dependence over m}
	{\it For any $ n $, $ S_n[m] $ and $ W_n[m] $ are independent over $ m $, while $ Z_n[m] $ is not and satisfies
		\begin{align} \label{eq: C(ZZ)}
			\myCC{Z_n[m_1],Z_n[m_2]}={\frac{\sin\left( \frac{2\pi}{\tilde{M}} \frac{\tilde{Q}(m_1-m_2)}{2} \right)}{{\tilde{Q}\sin\left( \frac{2\pi}{\tilde{M}} \frac{(m_1-m_2)}{2} \right)}}}.
		\end{align}
}
\end{lemma}

\mySpaceTwoMM

From 
(\ref{eq: V k r l our rdm}) and (\ref{eq: V k c l our rdm}), 
we notice that the SINR improvement is maximized when the IN background is independent over the range-Doppler grids, also known as `white'. However, we see from Lemmas \ref{lm: Sn[m] Wn[m] Zn[m] dependence} and \ref{lm: Zn[m] dependence over m}, the interference and noise signals are somewhat dependent over range-Doppler grids. More interestingly, there is a trade off in this regard caused by $ \frac{\tilde{Q}}{\tilde{M}} $. To reduce the correlation of $ S_n[m] $ and $ W_n[m] $ along $ n $, we prefer $ \tilde{M}\gg (\tilde{Q}+\bar{Q}) $ which also means $  \tilde{M}\gg \tilde{Q}$. However, according to (\ref{eq: C(ZZ)}), reducing $ \frac{\tilde{Q}}{\tilde{M}} $ will heavily increase the correlation of $ Z_n[m] $ over $ m $. In an extreme case, consider $ \tilde{Q} $ takes one, the smallest value. We then have $ \myCC{Z_n[m_1],Z_n[m_2]}=1~(\forall m_1,m_2) $. 
{As will be shown shortly, the dependence of $ S_n[m] $, $ W_n[m] $ and $ Z_n[m] $ over $ n $ and $ m $ makes it difficult to analyze the distribution of the IN background in the RDMs.} This, nevertheless, will be conquered.

\subsection{Analyzing Signal Components in Two RDMs}
\label{subsec: analyze signal components in RDMs}
We start with analyzing the distribution of the IN background, i.e., $ {W}_k^{\mathrm{r}}[l] + Z_k^{\mathrm{r}}[l] $, in the ratio-based RDM. According to (\ref{eq: V k r l our rdm}), $ {W}_k^{\mathrm{r}}[l] + Z_k^{\mathrm{r}}[l] $ can be rewritten as
\begin{align} \label{eq: Z kr[l]+W kr[l] computation}
	& {Z}_k^{\mathrm{r}}[l] + {W}_k^{\mathrm{r}}[l] = \sum_{n=0}^{\tilde{N}-1}\sum_{m=0}^{\tilde{M}-1} \frac{\overbrace{\left(Z_n[m] + {W}_n[m]\right) \myDFT{\tilde{M}}{-ml}\myDFT{\tilde{N}}{nk}}^{D_{n,m}^{k,l}}}{{S}_n[m]}.
\end{align}
Since $ Z_n[m] $ and ${W}_n[m] $ are independent Gaussian variables, their sum is also Gaussian distributed. Moreover, given $ \forall k,m,n,l $, $ \myDFT{\bar{M}}{-lm}$ and $\myDFT{\tilde{N}}{nk} $ have deterministic values, as defined in (\ref{eq: DFT basis}). Accordingly, applying Lemma \ref{lm: Sn[m] Wn[m] Zn[m] distributions}, we obtain 
\begin{align}\label{eq: D nm kl CN(0,)}
	D_{n,m}^{k,l}\sim\mathcal{CN}\left(0,(\sigma_Z^2+\sigma_W^2)\big/\tilde{M}\tilde{N}\right),
\end{align}
where the coefficient of the variance is from the two DFT bases; see (\ref{eq: DFT basis}).
Then, 
the summand in (\ref{eq: Z kr[l]+W kr[l] computation}) becomes the ratio of two uncorrelated complex Gaussian variables. Such a ratio conforms to a Cauchy distribution \cite{ComplexGaussianRatio_2010Globecom}.  
Now that $ {Z}_k^{\mathrm{r}}[l] + {W}_k^{\mathrm{r}}[l] $ becomes the sum of Cauchy variables, one would think of using the central limit theorem (CLT) to approximate the summation as a Gaussian distribution. Unfortunately, CLT is not applicable to Cauchy variables, as they have infinite variances \cite{ComplexGaussianRatio_2010Globecom}.
To this end, we provide a remedy below, in light of the fact that the CLT is applicable to the truncated Cauchy distributions \cite{Cauchy_truncated_hampel1998statistics}.

Instead of dividing $ S_n[m] $ directly, we can divide $ \ma S_n[m] $ with a real positive coefficient $ \ma $. 
Since $ \ma S_n[m]\sim\myCN{0,\ma^2\sigma_d^2} $ according to Lemma \ref{lm: Sn[m] Wn[m] Zn[m] distributions}, we can take a sufficiently large $ \ma $ such that the probability of the event $ |\ma S_n[m]|<1 $
can be reduced to a small value, say $ \epsilon $.
Moreover, if $ \epsilon I <1  $, then out of $ I $ samples of $ \myCN{0,\ma^2\sigma_d^2}  $, the event $ |\ma S_n[m]|<1 $ may not happen at all. According to \cite[Lemma 3]{Kai_otfs_IoTindustrial},
the critical value of $ \ma $, leading to $ \epsilon=1/I $, can be given by 
\begin{align}
	\ma_{\mathrm{c}} = 1\Big/\left( \sigma_d\sqrt{\ln \frac{I-1}{I}} \right).
\end{align}
Based on the above illustration, we can revise the ratio-based RDM as follows,
	\begin{align} \label{eq: V k r l our rdm revised}
	& \tilde{V}_k^{\mathrm{r}}[l] = \sum_{n=0}^{N-1}\sum_{m=0}^{M-1}\mathbb{I}_{\mathcal{E}}\left\{\frac{X_n[m]}{\ma S_n[m]}\right\} \myDFT{M}{-ml}\myDFT{N}{nk} \nonumber\\
	& \approx \sum_{p=0}^{P-1}\frac{\tilde{\alpha}_p}{\ma} \mySinc{\bar{M}}{l-\ml_p} \mySinc{\tilde{N}}{\mk_p-k} + \tilde{W}_k^{\mathrm{r}}[l] + \tilde{Z}_k^{\mathrm{r}}[l],\nonumber\\
	&\mathrm{s.t.}~\tilde{\mX}_k^{\mathrm{r}}[l] = \sum_{n=0}^{N-1}\sum_{m=0}^{M-1} \mathbb{I}_{\mathcal{E}}\left\{ \frac{\mX_n[m]}{\ma S_n[m]} \right\} \myDFT{M}{-ml}\myDFT{N}{nk},\mX\in\{W,Z\}\nonumber\\
	&~~~~~~ \mathbb{I}_{\mathcal{E}}\{\cdot\}=1~\text{if event $ \mathcal{E} $ happens; otherwise } \mathbb{I}_{\mathcal{E}}\{\cdot\}=0,\nonumber\\
	&~~~~~~ \mathcal{E}\myDef\{|\ma S_n[m]|\ge1\},
\end{align}
where $ X_n[m] $ is given in (\ref{eq: X n[m] new sensing}) and $ \ma S_n[m] $ is used as the divisor
compared with using $ S_n[m] $ in (\ref{eq: V k r l our rdm}). 
Note that the approximation is based on that $ \mathbb{I}_{\mathcal{E}}=0 $ can barely happen with a sufficiently large $ \ma $. 
For the same reason, we will drop the operator $ \mathbb{I}_{\mathcal{E}}\{\cdot\} $ below for notation simplicity. But bear in mind that $ \mathbb{I}_{\mathcal{E}}\{\cdot\} $ shall still be applied. 

Now we are able to invoke the CLT based on (\ref{eq: V k r l our rdm revised}). 
However, there is one more trap --- the summands under the CLT need to be i.i.d., while, as indicated by Lemmas \ref{lm: Sn[m] Wn[m] Zn[m] dependence} and \ref{lm: Zn[m] dependence over m}, the i.i.d. condition is not satisfied here. To this end, we resort to the case $ \tilde{M}\gg (\tilde{Q}+\bar{Q}) $, under which the correlation of $ S_n[m] $ and $ W_n[m] $ over $ n $ can be negligibly weak. As illustrated at the end of Section \ref{subsec: preliminary results}, $ \tilde{M}\gg (\tilde{Q}+\bar{Q}) $ can severely increase the correlation of $ Z_n[m] $ over $ m $. Nevertheless, we discover that applying the CLT along the $ n $-dimension first can approximately remove the correlation of the resulted Gaussian distributions over $ m $. More details are given in Appendix \ref{app: proof of proposition on IN of ratio-RDM}, while the results are summarized in the following proposition.

\mySpaceTwoMM

\begin{proposition}\label{pp: interference and noise of ratio-RDM}
	{\it Provided that $ \tilde{N} $ is large and $ \tilde{M}\gg (\tilde{Q}+\bar{Q}) $, the IN background of the ratio-based RDM obtained in (\ref{eq: V k r l our rdm revised}) approaches a complex centered Gaussian distribution which satisfies
	\begin{align} \label{eq: W r Z r CN(0,...)}
		&~~~~\tilde{W}_k^{\mathrm{r}}[l] + \tilde{Z}_k^{\mathrm{r}}[l]  \sim \myCN{0,\frac{(\sigma_Z^2+\sigma_W^2)\mb(\epsilon)}{\ma^2\sigma_d^2}},\\
		&\mathrm{s.t.}~
\mb(\epsilon) = 2\ln\left( {2(1-\epsilon)}\Big/{\left(\me\sqrt{\epsilon(2-\epsilon)}\right)} \right),
\nonumber	
\end{align}	
where $ \sigma_Z^2 $ and $ \sigma_W^2 $ are given in Lemma \ref{lm: Sn[m] Wn[m] Zn[m] distributions}, $ \epsilon $ is a sufficiently small number and $ \me $ denotes the base of the natural logarithm.
}
\end{proposition}

\mySpaceTwoMM

Despite the complex expression of the variance in (\ref{eq: W r Z r CN(0,...)}), it actually has a clear structure. Specifically, the fraction 
$ \frac{(\sigma_Z^2+\sigma_W^2)}{\ma^2\sigma_d^2} $ is the ratio between the variance of $ D_{n,m}^{k,l} $ and that of $ \ma S_n[m] $; in parallel with that $ \tilde{W}_k^{\mathrm{r}}[l] + \tilde{Z}_k^{\mathrm{r}}[l]  $
is the ratio between the two random variables. 
Such a ratio is known to have a heavy-tail PDF \cite{ComplexGaussianRatio_2010Globecom} and hence the coefficient $ \mb(\epsilon) $, greater than one in general, acts like a penalty factor to account for the heavy tail. 
Next, we elaborate more on $ \mb(\epsilon) $.
 According to \cite[Appendix D]{Kai_otfs_IoTindustrial}, $ \epsilon $ is the probability that $ \left|\Re\{\tilde{W}_k^{\mathrm{r}}[l] + \tilde{Z}_k^{\mathrm{r}}[l]\}\right| $ is larger than a threshold, where $ \Re\{\} $ takes the real part of a complex number. 
Regardless of the specific expression of the threshold, we hope $ \epsilon $ is such a small probability that out of 
$ \tilde{M}\tilde{N} $ samples of $ \tilde{W}_k^{\mathrm{r}}[l] + \tilde{Z}_k^{\mathrm{r}}[l] $, less than one sample can have the magnitude of its real part exceed the threshold.
So a critical value of $ \epsilon $ is $ 1/\big( \tilde{M}\tilde{N} \big) $. Substituting the value into (\ref{eq: W r Z r CN(0,...)}) leads to
\begin{align}\label{eq: bc}
	\mb_{\mathrm{c}} = \mb\left( 1/\big( \tilde{M}\tilde{N} \big) \right) = 2\left(\ln\left( \frac{2(\tilde{M}\tilde{N}-1)}{\sqrt{2\tilde{M}\tilde{N}-1}} \right) - 1 \right).
\end{align}
Note that $ \tilde{M}\tilde{N}\approx I $, the number of samples in the whole block can be tens of thousands and even greater.

With reference to the analysis yielding Proposition \ref{pp: interference and noise of ratio-RDM}, we can similarly analyze the distributions of the signal components in the CCC-based RDM obtained in (\ref{eq: V k c l our rdm}). This time, the CLT is directly applicable to the summands in (\ref{eq: V k c l our rdm}), as each is a product of two Gaussian variables and has a limited variance \cite{Gaussian_productOfcomplexGaussians}. 
However, the useful signal in the CCC-based RDM is substantially different from that in the ratio-based RDM; see (\ref{eq: V k c l our rdm}) and (\ref{eq: V k r l our rdm revised}). Thus, we provide some more analysis in Appendix \ref{app: proof of proposition on gaussian interfernce plus noise CCC}, with the focus on the useful signal in the CCC-based RDM. 
The following proposition summarizes the analytical results.

\mySpaceTwoMM

\begin{proposition}\label{pp: signal components in RDM ccc}
	{\it
		Provided large $ \tilde{M} $ and $ \tilde{N} $ as well as $ \tilde{M}\gg (\tilde{Q}+\bar{Q}) $, the signal components of the RDM obtained in (\ref{eq: V k c l our rdm}) approach Gaussian distributions that approximately satisfy
		\begin{align}\label{eq: Skc[l] CN, Wkcl+Zkcl CN}
			&S_k^{\mathrm{c}}[l]\sim\left\{
			\begin{array}{ll}
				\myN{\sigma_d^2\sqrt{\tilde{M}\tilde{N}},\sigma_d^4} & l= \ml_p~\mathrm{and}~k=\mk_p\\
				\myN{0,\sigma_d^4} & l\ne \ml_p~\mathrm{or}~k\ne \mk_p
			\end{array}
			\right.;\\
			&~~~~~~~{W}_k^{\mathrm{c}}[l] + Z_k^{\mathrm{c}}[l]\sim \mathcal{CN}\Big( 0,\sigma_d^2(\sigma_Z^2+\sigma_W^2) \Big).
		\end{align}
	}
\end{proposition}

{Detecting in Gaussian IN background has been widely studied for decades. Thus, 
	Propositions \ref{pp: interference and noise of ratio-RDM} and \ref{pp: signal components in RDM ccc} allow us to employ many existing detectors to detect targets from the RDMs obtained under the proposed sensing framework. The CA-CFAR has been briefly reviewed in Section \ref{sec: sensing framework}.
The two propositions also allow us to analyze and compare the SINRs in the two RDMs and draw insights into sensing parameter design. This is carried out next.  
}

\subsection{Comparison and Insights} \label{subsec: comparision and insights}

{The SINRs in the two RDMs are first derived based on Propositions \ref{pp: interference and noise of ratio-RDM} and \ref{pp: signal components in RDM ccc}.
}
Based on (\ref{eq: V k r l our rdm revised}), the power of the useful signal in the ratio-based RDM can be given by $ \sigma_P^2 \tilde{M}\tilde{N}/\ma^2 $, where $ \sigma_P^2=\sum_{p=0}^{P-1}\sigma_p^2 $. Then, combining the power of the IN terms derived in Proposition \ref{pp: interference and noise of ratio-RDM}, we obtain the SINR of the ratio-based RDM, as given by
\begin{align}\label{eq: gamma_r V}
	& \gamma_V^{\mathrm{r}} = \frac{ \tilde{M}\tilde{N}\sigma_P^2 \sigma_d^2 }{ (\sigma_Z^2+\sigma_W^2)\mb(\epsilon) } = \frac{ \tilde{M}\tilde{N} }{ \left(
		\frac{{\tilde{Q}}}{\tilde{M}} + 
		\left( 1 + \frac{\tilde{Q}}{\tilde{M}} \right)\frac{1}{\gamma_0\sigma_P^2}
		\right)\mb(\epsilon) }, \nonumber\\
&	\mathrm{s.t.}~\gamma_0 = {\sigma_d^2}\Big/{\sigma_w^2},~\sigma_P^2=\sum_{p=0}^{P-1}\sigma_p^2,
\end{align}
 where the expressions of $ \sigma_Z^2 $ and $ \sigma_W^2 $ given in Lemma \ref{lm: Sn[m] Wn[m] Zn[m] distributions} are used to get the final result.
 The above SINR can be simplified under certain asymptotic conditions. In particular, we have 
 \begin{align} \label{eq: gamma_V r gamma_0 limit}
 	&\gamma_V^{\mathrm{r}}  \left\{
 		\begin{array}{l}
				\myApprOverset{\gamma_0\ll \frac{1}{\sigma_P^2}}   \frac{\tilde{M}\left(  \frac{(I-\tilde{Q}-\bar{Q})}{(\tilde{M}-\bar{Q})} \right) \gamma_0\sigma_P^2 }{ \left( 1 + \frac{\tilde{Q}}{\tilde{M}} \right) \mb(\epsilon)  } \myApprOverset{(a)} \frac{I\gamma_0\sigma_P^2}{\left( 1 - \frac{\bar{Q}}{\tilde{M}} \right) \left( 1 + \frac{\tilde{Q}}{\tilde{M}} \right)  \mb(\epsilon)};\\
				\myApprOverset{\gamma_0\gg  \frac{1}{\sigma_P^2}} {I}\left/{\left(\left( 1 - \frac{\bar{Q}}{\tilde{M}} \right) \frac{\tilde{Q}}{\tilde{M}}  \mb(\epsilon)\right)}\right.
 		\end{array}
 	\right. \nonumber\\[-5mm] 	
 \end{align}
 where $ \myApprOverset{\gamma_0\ll \frac{1}{\sigma_P^2}} $ is obtained by (I) suppressing $ \frac{{\tilde{Q}}}{\tilde{M}} $ from the denominator of (\ref{eq: gamma_r V}) as $ \gamma_0\ll \frac{1}{\sigma_P^2} $ leads to $ \frac{{\tilde{Q}}}{\tilde{M}} \ll 
 \left( 1 + \frac{\tilde{Q}}{\tilde{M}} \right)\frac{1}{\gamma_0\sigma_P^2} $; and (II) replacing $ \tilde{N} $ with its expression given in (\ref{eq: s n[l]}) while suppressing the flooring operator. Moreover, $ \myApprOverset{(a)} $ is due to $ I=(M+Q)N\gg (\tilde{Q}+\bar{Q}) $. The second line in (\ref{eq: gamma_V r gamma_0 limit}) can be obtained similarly.

 For the CCC-based RDM, its SINR can be obtained by applying Proposition \ref{pp: signal components in RDM ccc} in (\ref{eq: V k c l our rdm}). 
 In particular, we have 
\begin{align}\label{eq: gamma_c V}
	\gamma_V^{\mathrm{c}} = \frac{ (\tilde{M}\tilde{N} + 1 )\sigma_P^2\sigma_d^4 }{ (\sigma_Z^2+\sigma_W^2)\sigma_d^2 + \sigma_P^2\sigma_d^4 } = \frac{ (\tilde{M}\tilde{N} + 1 ) }{ \frac{{\tilde{Q}}}{\tilde{M}} + 
		\left( 1 + \frac{\tilde{Q}}{\tilde{M}} \right)\frac{1}{\gamma_0\sigma_P^2} + 1 },
\end{align} 
where $ \sigma_P^2\sigma_d^4 $ in the denominator of the middle result is the interference caused by 
$ S_k^{\mathrm{c}}[l] $ at $ l\ne \ml_p $ or $ k\ne \mk_p $. 
With reference to the way (\ref{eq: gamma_V r gamma_0 limit}) is obtained, we can also attain the asymptotic $ \gamma_V^{\mathrm{c}} $, as given by
\begin{align}\label{eq: gamma_c V gamma_0 limit}
	 \gamma_V^{\mathrm{c}}  \left\{
	 	\begin{array}{l}
	 		\myApprOverset{\gamma_0\ll \frac{1}{\sigma_P^2}}   {I\gamma_0\sigma_P^2}\left/{\left(\left( 1 - \frac{\bar{Q}}{\tilde{M}} \right) \left( 1 + \frac{\tilde{Q}}{\tilde{M}} \right)\right) }\right.\\
	 		\myApprOverset{\gamma_0\gg  \frac{1}{\sigma_P^2}} {I}\left/{\left(\left( 1 - \frac{\bar{Q}}{\tilde{M}} \right) \left( 1 + \frac{\tilde{Q}}{\tilde{M}} \right) \right)
	 			}\right.
	 	\end{array}
	 \right..
\end{align}

The SINRs derived in (\ref{eq: gamma_r V}) and (\ref{eq: gamma_c V}) can be adapted for the RDMs obtained in the framework of COS, i.e., 
(\ref{eq: U r k[l]}) and (\ref{eq: U c k[l]}). As reviewed in Section \ref{subsec: reviewing classical ofdm sensing}, COS complies with the underlying communication system. 
Thus, we can take $ \tilde{M}=M $ and $ \tilde{N}=N $ in (\ref{eq: gamma_r V}) and (\ref{eq: gamma_c V}). Moreover, since COS uses the original communication CP, $ \sigma_Z^2 $ in (\ref{eq: gamma_r V}) and (\ref{eq: gamma_c V}), which is the power of the interference caused by the proposed VCP, can be suppressed, and $ \sigma_W^2$ can be replaced by $\sigma_w^2 $. Under the above changes, the SINRs of the RDMs in (\ref{eq: U r k[l]}) and (\ref{eq: U c k[l]}) can be, respectively, given by
\begin{equation}
	\begin{gathered}\label{eq: gamma_U r and gamma_U c}
		\gamma_U^{\mathrm{r}} = \frac{MN\gamma_0\sigma_P^2}{\mb(\epsilon)}
		\myEqualOverset{(a)} \frac{I\gamma_0\sigma_P^2}{\left(1+\frac{Q}{M}\right) \mb(\epsilon) }~(\forall \gamma_0);\\
		\gamma_U^{\mathrm{c}} =\frac{(MN+1)\gamma_0\sigma_P^2}{1+\gamma_0\sigma_P^2}\left\{ \begin{array}{l}
			\myApprOverset{\gamma_0\ll \frac{1}{\sigma_P^2}} {I\gamma_0\sigma_P^2}\Big/{\left(1+\frac{Q}{M}\right) } \\
			\myApprOverset{\gamma_0\gg \frac{1}{\sigma_P^2}} {I}\Big/{\left(1+\frac{Q}{M}\right)  }
		\end{array} \right.  	
	\end{gathered}
\end{equation}
where $ \myEqualOverset{(a)} $ is obtained by replacing $ N $ with $ \frac{I}{(M+Q)} $, the same replacement is also performed for $ 	\gamma_U^{\mathrm{c}} $, and the approximations are similarly attained, as done in (\ref{eq: gamma_V r gamma_0 limit}). 
Now, we are ready to make some comparisons using the SINR expressions.

\mySpaceTwoMM

\begin{remark}\label{rmk: ratio based rdm ours versus COS}
		For the ratio-based RDM, we make the following comparisons between COS and the proposed sensing:
	\begin{enumerate}[leftmargin=6mm]\renewcommand{\labelenumi}{{\ref{rmk: ratio based rdm ours versus COS}\alph{enumi})}}
		\item {In low SNR regions, such that $ \gamma_0 =\frac{\sigma_d^2}{\sigma_w^2}  \ll \frac{1}{\sigma_P^2} $, the proposed sensing framework has a greater SINR than COS with a gain no less than $ \frac{1}{1-\frac{\bar{Q}}{\tilde{M}}} $, provided $ \frac{\tilde{Q}}{\tilde{M}}\le \frac{Q}{M} $;}

		\item Provided the maximum round-trip delay of a target is smaller than the communication CP duration, i.e., 		
		$ \max_{\forall p}\{\tau_p\}\le QT_{\mathrm{s}} $, COS can have a greater SINR than the proposed sensing framework for $ \gamma_0>\frac{1+\frac{Q}{M}}{\left( 1-\frac{\bar{Q}}{\tilde{M}}  \right)\frac{\tilde{Q}}{\tilde{M}} \sigma_P^2} $;
		
		\item Provided $ \max_{\forall p}\{\tau_p\}> QT_{\mathrm{s}} $, 
		the result in Remark \ref{rmk: ratio based rdm ours versus COS}b) may not hold any more; moreover, the proposed sensing framework can have a greater SINR than COS;
		
	\end{enumerate}	
	The first two results can be readily attained based on (\ref{eq: gamma_V r gamma_0 limit}) and (\ref{eq: gamma_U r and gamma_U c}). It is noteworthy that the condition 
$ \max_{\forall p}\{\tau_p\}\le QT_{\mathrm{s}} $ is implicitly required by COS to remove communication data symbols for generating RDMs; see the review in Section \ref{subsec: reviewing classical ofdm sensing}. 
If the condition is unsatisfied, the SINR of COS, as given in (\ref{eq: gamma_U r and gamma_U c}), becomes invalid. 
However, for the fact that COS cannot effectively remove communication data symbols any more while the proposed sensing can, we attain the result in Remark \ref{rmk: ratio based rdm ours versus COS}c).
As will be validated by Fig. \ref{fig: sinr2 Qtilde} in Section \ref{sec: simulations}, the SINRs of the two RDMs under COS degrade severely, as $ \max_{\forall p}\{\tau_p\} $ exceeds $ QT_{\mathrm{s}} $. 
\end{remark}
\renewcommand{\labelenumi}{{\arabic{enumi})}}

\mySpaceTwoMM

\begin{remark}\label{rmk: ccc based rdm ours versus COS}		
		For the CCC-based RDM, provided $ \frac{\tilde{Q}}{\tilde{M}}\le \frac{Q}{M} $, the proposed sensing always has a greater SINR than COS regardless of $ \gamma_0 $ and the SINR gain is no less than $ \frac{1}{1-\frac{\bar{Q}}{\tilde{M}}} $.
The results can be easily validated using (\ref{eq: gamma_c V gamma_0 limit}) and (\ref{eq: gamma_U r and gamma_U c}). 
{An intriguing question is why the relationship between COS and the proposed sensing is substantially different under the two RDMs. In essence, this is caused by the different ways the communication data symbols are removed for generating the two RDMs. For the ratio-based RDM, 
the pointwise division, as shown in (\ref{eq: V k r l our rdm}), magnifies the IN background by introducing the multiplicative coefficient $ \mb(\epsilon) $. 
In contrast, the pointwise product for the CCC-based RDM; see (\ref{eq: V k c l our rdm}), introduces an additive interference $ \sigma_P^2\sigma_d^4 $; see (\ref{eq: gamma_c V}).
While the VCP-incurred interference does not bother COS, the CCC-incurred interference exists in both COS and the proposed sensing framework. 
}

\end{remark}
\renewcommand{\labelenumi}{{\arabic{enumi})}}

\mySpaceTwoMM

\begin{remark} \label{rmk: ratio versus ccc rdms}
Some comparisons between the ratio- and CCC-based RDMs are made here. Based on (\ref{eq: gamma_V r gamma_0 limit}) and (\ref{eq: gamma_c V gamma_0 limit}), we can attain the following results for 
the proposed sensing: 
\begin{enumerate}[leftmargin=6.5mm]\renewcommand{\labelenumi}{{\ref{rmk: ratio versus ccc rdms}\alph{enumi})}}
	\item In low SNR regions where $ \gamma_0\ll 1/\sigma_P^2 $, the CCC-based 
	RDM has an SINR that is $ \mb(\epsilon) $ times the SINR in the ratio-based RDM, where $ \mb(\epsilon)>1 $ in general; see (\ref{eq: bc}). 
	
	\item In high SNR regions where $ \gamma_0\gg 1/\sigma_P^2 $, the ratio-based RDM can have a greater SINR than the CCC-based RDM, provided $ \mb(\epsilon)\le \frac{\tilde{M}}{\tilde{Q}} $. 
	
	\item Regardless of $ \gamma_0 $, the CCC-based RDM always has a greater SINR than the ratio-based RDM, if $ \mb(\epsilon)>\frac{\tilde{M}}{\tilde{Q}}+1 $. 
	
	\newcounter{nameOfYourChoice}
	\setcounter{nameOfYourChoice}{\value{enumi}}
\end{enumerate}
Based on (\ref{eq: gamma_U r and gamma_U c}), similar results as above can be given for COS:
\begin{enumerate}\setcounter{enumi}{\value{nameOfYourChoice}}\renewcommand{\labelenumi}{{\ref{rmk: ratio versus ccc rdms}\alph{enumi})}}
	\item The result in \ref{rmk: ratio versus ccc rdms}a) directly applies to COS;
	
	\item In high SNR regions where $ \gamma_0\gg 1/\sigma_P^2 $, the CCC-based RDM has a greater SINR than the ratio-based RDM, if $ \mb(\epsilon)>\gamma_0\sigma_P^2 $, while if $ \mb(\epsilon)<\gamma_0\sigma_P^2 $ the ratio-based RDM has a greater SINR. 
\end{enumerate}

\end{remark}
\renewcommand{\labelenumi}{{\arabic{enumi})}}

\subsection{Criteria for Setting Key Sensing Parameters}
\label{subsec: adjust M N for our design}
Unlike COS that follows with the underlying communication system, our sensing framework has the flexibility of catering different sensing needs via adjusting several key parameters: $ \tilde{M}  $, $ \tilde{Q} $ and $ \bar{Q} $ ($ \tilde{N} $ is determined given the former three). Below, we illustrate 
the criteria for setting these parameters to optimize sensing performance.

\textit{First}, we can set $ \tilde{Q} $ based on the required maximum sensing distance, as denoted by $ r_{\mathrm{max}} $.
From Section \ref{sec: sensing framework}, the sensing distance of the proposed design is given by $ \frac{\mC\tilde{Q}\ts}{2} $, which, equating with $ r_{\mathrm{max}} $, yields $ \tilde{Q}=\frac{2r_{\mathrm{max}}}{\mC\ts} $. {It is worth noting that the issue of CP-limited sensing, as described in Section \ref{subsubsec: CP limitation}, is addressed by introducing $ \tilde{Q} $. Unlike in COS and its variants where $ r_{\mathrm{max}} $ is determined by $ Q $, we now can set $ \tilde{Q} $ to satisfy $ r_{\mathrm{max}} $ (provided a sufficient link budget).}

\textit{Second}, we determine $ \tilde{M} $ given the requirements on velocity measurement.
Applying the analysis in Section \ref{subsubsec: M and N selection}, the maximum measurable value and the resolution of Doppler frequency of the proposed sensing framework are given by
\[
\nu_{\mathrm{max}} = {1}\Big/\big(2(\tilde{M}-\bar{Q})T_{\mathrm{s}}\big);~
\dnu \approx {1}\Big/\big(IT_{\mathrm{s}}\big). \nonumber
\]
Thus, to cater the expected $ \nu_{\mathrm{max}}^* $ 
	we need to keep $ \tilde{M}\le {1}\Big/\big(2\nu_{\mathrm{max}}^*T_{\mathrm{s}}\big)+\bar{Q} $. Moreover, we prefer to have a relatively large\footnote{
		Note that, if $ \bar{Q}/\tilde{M} $ is fixed, the larger $ \tilde{M} $ the smaller $ \tilde{N}(\approx I\big/(\tilde{M}(1-\bar{Q}/\tilde{M}))) $ will become. This may affect the precision of results given in Propositions \ref{pp: interference and noise of ratio-RDM} and \ref{pp: signal components in RDM ccc} which expect both $ \tilde{M} $ and $ \tilde{N} $.
	} $ \tilde{M} $ which can lead to a small $ \frac{\tilde{Q}}{\tilde{M}} $ and hence a high SINR in both RDMs; see (\ref{eq: gamma_V r gamma_0 limit}) and (\ref{eq: gamma_c V gamma_0 limit}). 
{It is noteworthy that the issue of limited velocity measurement, as illustrated in Section \ref{subsubsec: M and N selection}, is now addressed by introducing $ \tilde{M}$ and $\bar{Q} $. Instead of having an $ M $-limited $ \nu_{\mathrm{max}} $, we now have the flexibility of configuring $ \tilde{M} $ to satisfy the requirement on $ \nu_{\mathrm{max}} $.
}

\textit{Third},  
given $ \tilde{M} $, 
we can then set $ \bar{Q} $. {To increase the SINR in both RDMs, we expect to have $ \bar{Q} $ as large as possible; see (\ref{eq: gamma_V r gamma_0 limit}) and (\ref{eq: gamma_c V gamma_0 limit}). However, the larger $ \bar{Q} $ the more correlated the signals between adjacent sub-blocks 
	can be; see Lemmas \ref{lm: Sn[m] Wn[m] Zn[m] dependence} and \ref{lm: Zn[m] dependence over m}.} 
As seen from Propositions \ref{pp: interference and noise of ratio-RDM} and \ref{pp: signal components in RDM ccc}, the correlation can make the results less precise. 
The detailed impact, however, is difficult to analyze. As will be shown through the simulations in
Figs. \ref{fig: sinr2 Qbar ratio} and \ref{fig: sinr2 Qbar CCC}, the derivations and analysis in 
Sections \ref{subsec: analyze signal components in RDMs} and \ref{subsec: comparision and insights} are consistently precise when $ \bar{Q} $ takes from a small value to the one as large as $ \tilde{M}/2-\tilde{Q} $.

\section{Simulation Results}\label{sec: simulations}

\begin{table}[!t]\footnotesize
	\captionof{table}{Simulation Parameters}
	\vspace{-3mm}
	\begin{center}
		\begin{tabular}{m{0.05\linewidth}|m{0.42\linewidth}|m{0.35\linewidth}}
			\hline
			Var. &
			Description &
			IEEE 802.11ad 
			\\
			\hline
			$ f_{\mathrm{c}} $    
			&  Carrier frequency 
			&	$ 60.48 $ GHz 
			\\				
			\hline
			$ B $ 
			& Bandwidth 
			& $ 1.825 $ GHz
			\\				
			\hline
			$ M $ 
			& No. of sub-carriers per symbol
			& $ 512 $
			\\				
			\hline
			$ Q $ 
			& CP length
			& $ 128 $
			\\				
			\hline
			$ N $ 
			& No. of symbols
			& $ 143 $ (0.05 ms packet)
			\\				
			\hline
			$ I $ 
			& Total No. of samples; see (\ref{eq: s tilde CP [i]})
			& $ N(M+Q)=91 520 $ 
			\\				
			\hline	
			$ \sigma_d^2 $ 
			& Power of data symbol $ d_{i} $; see (\ref{eq: S[m,n]})
			& $ 0 $ dB
			\\				
			\hline
			$ \sigma_p^2 $ 
			& Power of $ \alpha_p $; see (\ref{eq: x[i]})
			& $ [0,-10,-20] $ dB
			\\				
			\hline
			
			$ r_p $ 
			& Target range
			& $ \mathcal{U}_{[0,10]~\mathrm{m}}~^{\dagger} $
			\\				
			\hline
			$ v_p $ 
			& Target velocity
			& $ \mathcal{U}_{[-139,139]~\mathrm{m/s}} $
			\\				
			\hline
			$ \sigma_w^2 $ 
			& Variance of AWGN $ w[i] $; see (\ref{eq: x[i]})
			& $ -20 $ dB
			\\				
			\hline
			
			\multicolumn{3}{l}{
				$ ^{\dagger} $ $ \mathcal{U}_{[x,y]} $ denotes a uniform distribution in the region give by the subscript.
			}
			\\
			\hline
			
		\end{tabular}
		\vspace{-3mm}
	\end{center}
	\label{tab: simulation parameters benchmark}
\end{table}	

Simulations are performed in this section to validate the proposed design. The simulation parameters are set with reference to \cite{DFRC_SC_OFDM} and are summarized in Table \ref{tab: simulation parameters benchmark}. 
The root raised cosine (RRC) filter with the roll-off coefficient of $ 0.2 $ is used at both the communication transmitter and the sensing receiver. In generating target echo signals, a four-times upsampling is performed by the transmitter RRC filter; the target delay and Doppler frequency are added at the high sampling rate; and a four-times decimating is performed at the receiver RRC filter. 
This generates off-grid range and Doppler values, making the simulations comply with practical scenarios.

The benchmark sensing framework is COS \cite{DFRC_dsss2011procIeee}, as reviewed in Section \ref{subsec: reviewing classical ofdm sensing}. 
The original COS, as developed for OFDM \cite{DFRC_dsss2011procIeee}, uses the ratio-based RDM given in (\ref{eq: U r k[l]}), while the variant of COS, as developed for \dso~\cite{DFRC_SC_OFDM}, employs the CCC-based RDM given in (\ref{eq: U c k[l]}). As illustrated in Remark \ref{rmk: otfs represents dft-s-ofdm and ofdm}, 
for both
OFDM and \dso,
the time-domain transmitted communication signals conform to Gaussian distributions, which is similar to OTFS. 
Therefore, for fair comparison, we unitedly use OTFS modulation for all methods to be simulated. 
In essence, it is the way a block of echo signal is segmented, rather than the communication waveforms, that differentiates COS and the proposed sensing framework.

\textit{In the legends of the simulation results, we use `r' to indicate the `ratio-based RDM', `c' the `CCC-based RDM', `sim' the simulated result and `pp' the `proposed design'.}

\begin{figure}[!t]
	\centering
	\includegraphics[width=80mm]{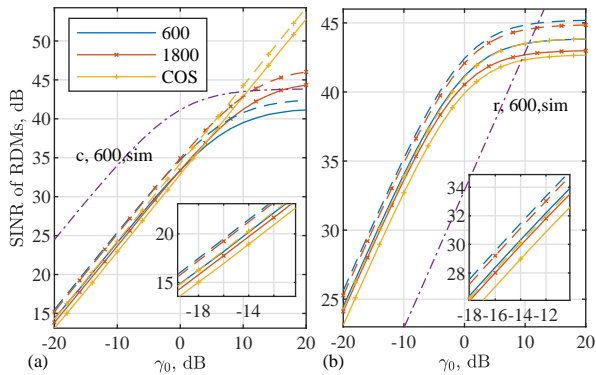}
	\caption{SINRs in the RDMs versus $ \gamma_0 $ defined in (\ref{eq: gamma_r V}). The ratio-based RDMs are shown in Fig. \ref{fig: sinr 2 snr diff Mtilde}(a), while the CCC-based RDMs in Fig. \ref{fig: sinr 2 snr diff Mtilde}(b). The two sub-figures share the same legend, where the numbers are the values of $ \tilde{M} $ used for the proposed design.
	Corresponding to the solid curves, the dash ones are the theoretical SINRs derived in (\ref{eq: gamma_r V}), (\ref{eq: gamma_c V}) and (\ref{eq: gamma_U r and gamma_U c}). For comparison convenience, the curve `c, 600, sim' in Fig. \ref{fig: sinr 2 snr diff Mtilde}(a) is copied from Fig. \ref{fig: sinr 2 snr diff Mtilde}(b) and the curve `r, 600, sim' in Fig. \ref{fig: sinr 2 snr diff Mtilde}(b) is from Fig. \ref{fig: sinr 2 snr diff Mtilde}(a).}
	\label{fig: sinr 2 snr diff Mtilde}
	
	\vspace{-5mm}
\end{figure}

\subsection{Illustrating SINRs in RDMs}

Fig. \ref{fig: sinr 2 snr diff Mtilde} plots the SINRs of the ratio- and CCC-based RDMs versus $ \gamma_0 $, under different values of $ \tilde{M} $. In this simulation, $ \tilde{Q}=Q $, $ \bar{Q} =150$ and other parameters are given in Table \ref{tab: simulation parameters benchmark}.
Overall, we see that the derived SINRs can precisely describe the actual (simulated) SINRs. 
This validates the analysis in Section \ref{subsec: analyze signal components in RDMs}.
More specifically,
we see from Fig. \ref{fig: sinr 2 snr diff Mtilde}(a) that the proposed design achieves higher SINRs in the ratio-based RDM than COS in the case of $ \gamma_0\ll 1 $, which validates 
Remark \ref{rmk: ratio based rdm ours versus COS}a). 
We also see Fig. \ref{fig: sinr 2 snr diff Mtilde}(a) that as $ \tilde{M} $ increases the gap between the proposed design and COS becomes smaller for $ \gamma\ll 1 $. This is consistent with the SINR expression derived in (\ref{eq: gamma_r V}). We further see that, when $ \gamma_0\gg 1 $, COS can outperform the proposed design, which complies with Remark \ref{rmk: ratio based rdm ours versus COS}b). We see from Fig. \ref{fig: sinr 2 snr diff Mtilde}(b) that the SINR achieved in the CCC-based RDM first increases with $ \gamma_0 $ and then converges for large $ \gamma_0 $'s. This is consistent with (\ref{eq: gamma_c V}). Moreover, we see that, for the CCC-based RDM,
the proposed design achieves the higher SINR across the whole region of $ \gamma_0 $ compared with COS. This aligns with Remark \ref{rmk: ccc based rdm ours versus COS}. 
A final note about Fig. \ref{fig: sinr 2 snr diff Mtilde} is that the comparison between the ratio- and CCC-based RDMs validates the analysis given in Remark \ref{rmk: ratio versus ccc rdms}.

\begin{figure}[!t]
	\centering
	\includegraphics[width=80mm]{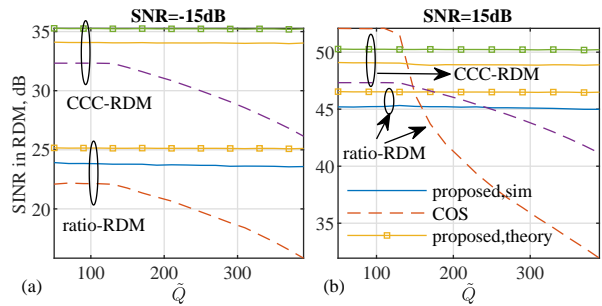}
	\caption{SINRs in RDMs versus $ \tilde{Q} $ (the length of VCP), where the values of $ \tilde{M} $ and $ \bar{Q} $ are changed with $ \tilde{Q} $ to keep $ \frac{\tilde{Q}}{\tilde{M}}=\frac{1}{4} $ and $ \frac{\bar{Q}}{\tilde{M}}=\frac{1}{3} $. A single target is simulated with its distance set as $ \frac{(\bar{Q}-1)\mC}{2B} $ m, while other parameters are set as in Table \ref{tab: simulation parameters benchmark}.}
	\label{fig: sinr2 Qtilde}
	\vspace{-5mm}
\end{figure}

Fig. \ref{fig: sinr2 Qtilde} illustrates a great flexibility of the proposed sensing framework by showing its achieved SINRs in the two types of RDMs under different values of $ \tilde{Q} $. 
As said in the caption of the figure, we keep the ratios $ \frac{\tilde{Q}}{\tilde{M}} $ and $ \frac{\bar{Q}}{\tilde{M}}$ fixed under different $ \tilde{Q} $'s. Then, according to (\ref{eq: gamma_r V}) and (\ref{eq: gamma_c V}), we know that the SINRs achieved by the proposed design should be the same over $ \tilde{Q} $, which is clearly validated by Fig. \ref{fig: sinr2 Qtilde}. In contrast, COS degrades severely when $ \tilde{Q} $ exceeds $ Q=128 $. This is because COS strictly follows the underlying communication system and cannot fully remove the communication data symbol when the echo delay is larger than the CP length; see (\ref{eq: Xn[m]}).

\begin{figure}[!t]
	\centering
	\includegraphics[width=80mm]{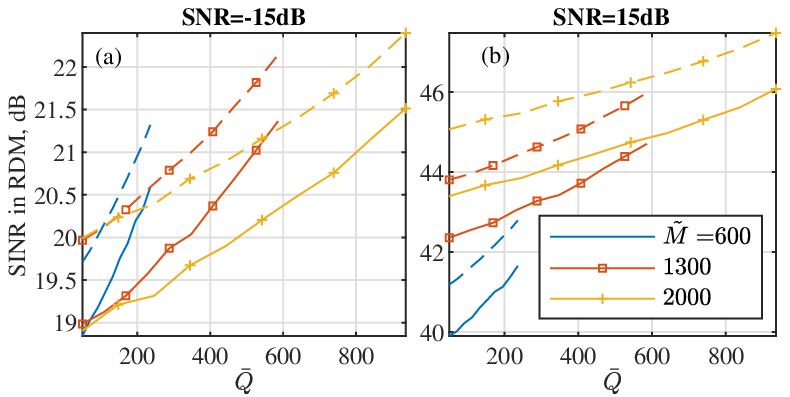}
	\caption{SINRs in the ratio-based RDM versus $ \bar{Q} $ (the number of overlapping samples between adjacent sub-blocks), where $ \tilde{Q}=Q $.}
	\label{fig: sinr2 Qbar ratio}
	
	\mySpaceTwoMM
	
	\includegraphics[width=80mm]{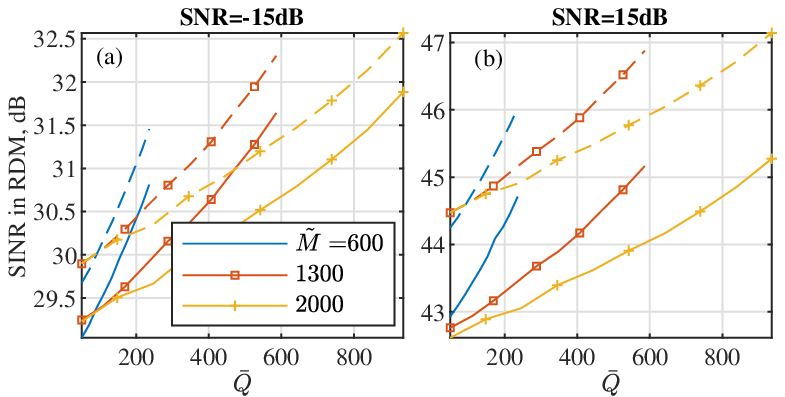}
	\caption{SINRs in the CCC-based RDM versus $ \bar{Q} $ (the number of overlapping samples between adjacent sub-blocks), where $ \tilde{Q}=Q $.}
	\label{fig: sinr2 Qbar CCC}
\end{figure}

Figs. \ref{fig: sinr2 Qbar ratio} and \ref{fig: sinr2 Qbar CCC} demonstrate another great flexibility of the proposed sensing framework by showing the SINRs achieved in the two types of RDMs under different values of $ \bar{Q} $. 
Overall, we see from the two figures that SINRs increase with $ \bar{Q} $ but the slopes decrease as $ \tilde{M} $ becomes larger. This can be well seen from the analytical SINR expressions derived in (\ref{eq: gamma_r V}) and (\ref{eq: gamma_c V}). Moreover, we see from Fig. \ref{fig: sinr2 Qbar ratio} that the SINR performance under low and high SNRs are different, while such phenomenon is not seen in Fig. \ref{fig: sinr2 Qbar CCC}. The rationale for this result can be seen from Remark \ref{rmk: ccc based rdm ours versus COS}. 
It is worth highlighting that, as seen from the figure, the analytical results match the simulated ones in the whole region of $ \bar{Q} $. This manifests the high flexibility of the proposed design.

\subsection{Illustration of Target Detection}
Next, we translate the SINR results obtained above into the actual detecting performance of the proposed sensing framework. To do so, we perform the 
CA-CFAR according to the steps given in Table \ref{tab: proposed sensing framework}. 
In the following simulations, the parameters in Step 6), Table \ref{tab: proposed sensing framework} are set as: $ N_{\mathrm{g}}^k= N_{\mathrm{g}}^l=3 $, $ N_{\mathrm{r}}^k=2$ and $  N_{\mathrm{r}}^l=5 $. 
For a better time efficiency of simulating and  calculating the detecting probability, we make two changes in the target scenario. \textit{First}, we increase the number of targets to $ P=10 $ and have their ranges linearly spaced in $ [0,10]  $ m. The velocities of the targets are still uniformly distributed, as illustrated in Table \ref{tab: simulation parameters benchmark}. \textit{Second}, the powers of the targets are set as follows: $ \sigma_0^2=0 $ dB, $ \sigma_p^2=-20 $ dB for $ p=1,\cdots,4 $ and $ \sigma_p^2=-30 $ dB for $ p=5,\cdots,9 $.

\begin{figure}[!t]
	\centering
	\includegraphics[width=80mm]{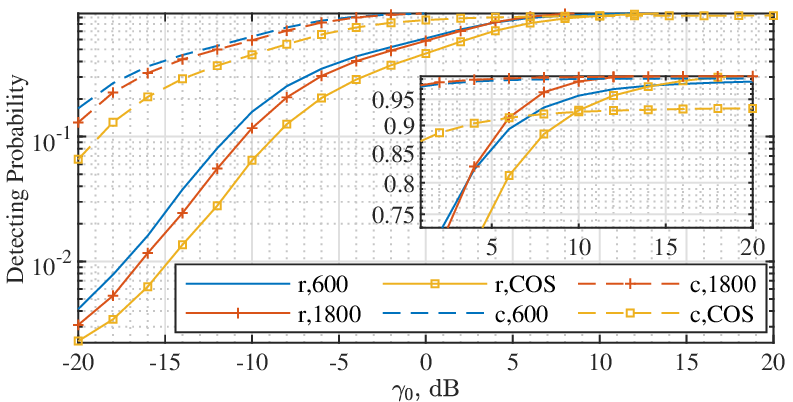}
	\caption{Illustration of the detection performance of the proposed sensing framework, where $ \tilde{Q}=Q $, $ \bar{Q}=150 $ and $ \mP_{\mathrm{F}}=10^{-6} $. The numbers in the legend are the values of $ \tilde{M} $.}
	\label{fig: pd2snr}
	
	\mySpaceTwoMM
	
	\includegraphics[width=80mm]{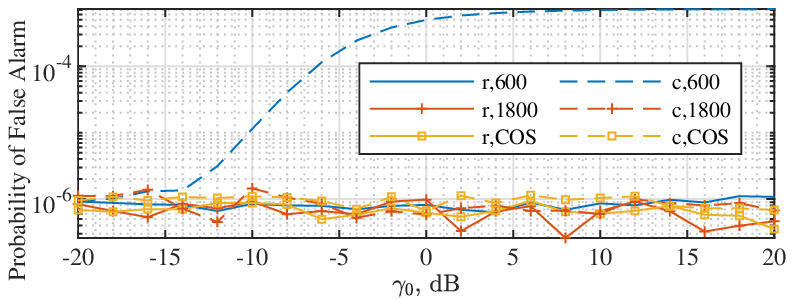}
	\caption{Illustrating $ \mP_{\mathrm{F}} $ corresponding to the curves in Fig. \ref{fig: pd2snr}.}
	\label{fig: pfa2snr}
	\vspace{-5mm}
\end{figure}

\begin{figure}
	\centering
	\includegraphics[width=80mm]{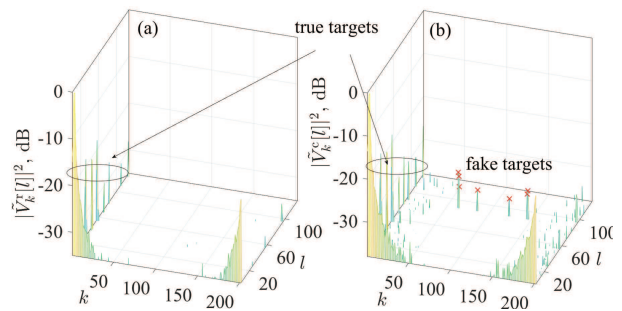}
	\caption{Comparing the ratio- and CCC-based RDMs, as given in Figs. \ref{fig: rdm4pfa explain}(a) and \ref{fig: rdm4pfa explain}(b), respectively. For illustration clarity, both RDMs are noise-less and the $ z $-axis is limited to avoid heavy interference background.}
	\label{fig: rdm4pfa explain}
	\vspace{-3mm}
\end{figure}

Fig. \ref{fig: pd2snr} shows the detecting probability of COS and the proposed sensing framework under different $ \tilde{M} $. 
We see from Fig. \ref{fig: pd2snr} that the proposed sensing framework achieves better detecting performance than COS for both the ratio- and CCC-based RDMs. We also see that the improvement of the detecting probability is precisely predicted by the SINR results observed in Fig. \ref{fig: sinr 2 snr diff Mtilde}. 
This not only manifests the superiority of the proposed design over COS in detecting performance but also validate our analysis and derivations provided in Section \ref{sec: analyze and compare RDMs}. Fig. \ref{fig: pfa2snr} shows $ \mP_{\mathrm{F}} $ versus $ \gamma_0 $ corresponding to each curve in Fig. \ref{fig: pd2snr}. We see that, as set, $ \mP_{\mathrm{F}}=10^{-6} $ is achieved in most cases. An exception, however, happens when using the CCC-based RDM under $ \tilde{M}=600 $. In such a case, $ \mP_{\mathrm{F}} $ increases with $ \gamma_0 $. 
The reason is that the overlapping of consecutive sub-blocks makes the essential signal of a block partially correlated with the interference from its previous sub-block; see Fig. \ref{fig: rearrange and VCP}. 
This is validated by Fig. \ref{fig: rdm4pfa explain}, where we can see the fake targets in the CCC-based RDM.

\begin{figure}
	\centering
	\includegraphics[width=80mm]{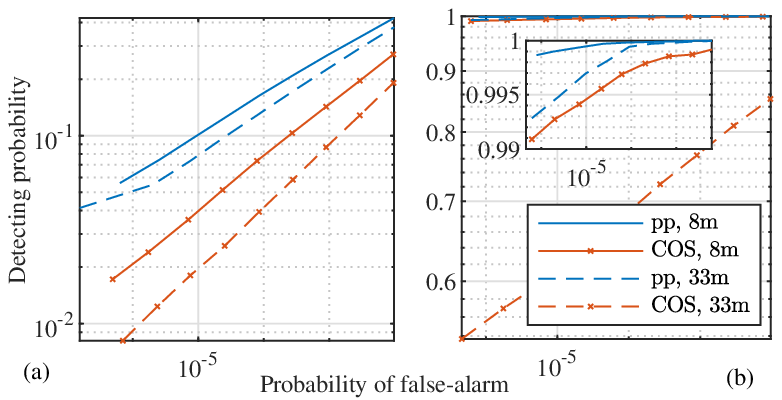}
	\caption{Comparing the receiver operating characteristic (ROC) of COS and the proposed (pp) sensing framework using the ratio-based RDM, where $ \gamma_0=-15 $ dB is set for Fig. \ref{fig: roc diff Qtilde ratio}(a) and $ \gamma_0=15 $ dB for Fig. \ref{fig: roc diff Qtilde ratio}(b). In the legend, $ 8 $ m and $ 33 $ m are the maximum ranges of targets.}
	\label{fig: roc diff Qtilde ratio}
	
	\mySpaceTwoMM
	
	\includegraphics[width=80mm]{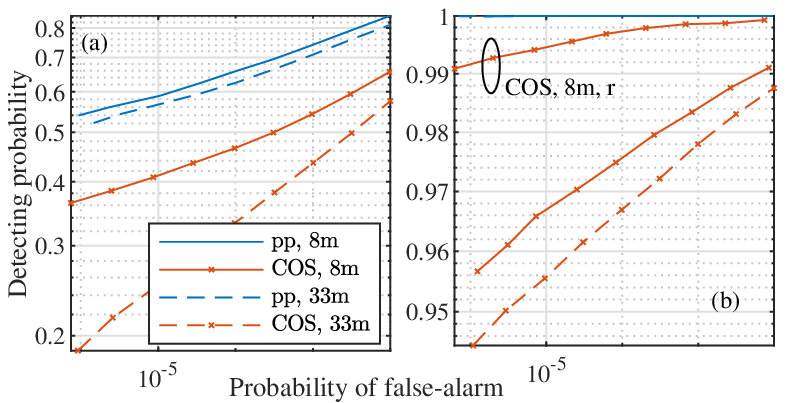}
	\caption{ROC curves under the CCC-based RDMs, corresponding to Fig. \ref{fig: roc diff Qtilde ratio}. The one labeled `COS, 8m, r' is copied from Fig. \ref{fig: roc diff Qtilde ratio}(b) for comparison.}
	\label{fig: roc diff Qtilde ccc}
	\vspace{-5mm}
\end{figure}

Figs. \ref{fig: roc diff Qtilde ratio} and \ref{fig: roc diff Qtilde ccc} illustrate the receiver operating characteristic (ROC) of the proposed sensing framework in comparison with that of COS. The cases of $ \tilde{Q}=100 $ and $ 400 $ in Fig. \ref{fig: sinr2 Qtilde} are 
considered here, corresponding to the maximum ranges of $ 8 $ m and $ 33 $ m, respectively.
From Figs. \ref{fig: roc diff Qtilde ratio} and \ref{fig: roc diff Qtilde ccc}, we see that the proposed design is robust under different maximum ranges, while COS, as predicted in Fig. \ref{fig: sinr2 Qtilde}, degrades severely when the maximum range exceeds that specified by underlying communication systems, i.e., $ 10 $ m. This demonstrates the superior flexibility of the proposed design in handling different sensing requirements.

\begin{figure}
	\centering
	\includegraphics[width=80mm]{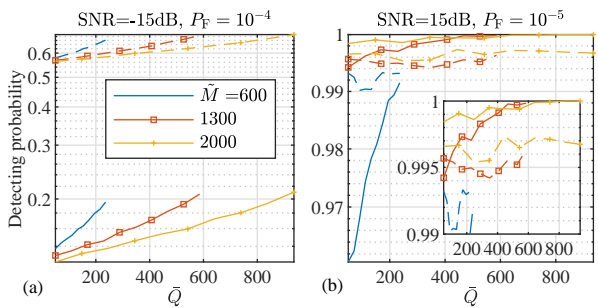}
	\caption{Illustration of the detecting probability versus $ \bar{Q} $ corresponding to the results presented in Figs. \ref{fig: sinr2 Qbar ratio} and \ref{fig: sinr2 Qbar CCC}.}
	\label{fig: pd2Qbar}
	\vspace{-5mm}
\end{figure}

Fig. \ref{fig: pd2Qbar} shows another flexibility of the proposed design from introducing $ \bar{Q} $. 
We see that, in overall, the detecting performance of the proposed design becomes better as $ \bar{Q} $ increases. This is consistent with the SINR results observed in Figs. \ref{fig: sinr2 Qbar ratio} and \ref{fig: sinr2 Qbar CCC}. We also see that for the low SNR shown in Fig. \ref{fig: pd2Qbar}(a), the impact of $ \bar{Q} $ is more prominent compared with that in the high SNR case shown in Fig. \ref{fig: pd2Qbar}(b). This is reasonable as $ \bar{Q} $ is introduced to increase the number of sub-blocks, and hence the SINR in the RDM. However, such improvement is limited as the larger $ \bar{Q} $, the higher correlation of the IN background over sub-blocks, as illustrated in Section \ref{subsec: preliminary results}.

\section{Conclusions}\label{sec:conclusions}

In this paper, we develop a novel sensing framework that is applicable to not only cyclic prefixed waveforms, e.g., OFDM and \dso, but also those with reduced CP, e.g., \rcpotfs. 
Unlike COS and its variants, we do not follow the underlying communication system and 
unprecedentedly achieves the flexibility of adapting for different sensing needs. This is achieved by a new block segmentation design that segments a whole block of echo signal evenly into multiple sub-blocks that can overlap between adjacent ones. This is also accomplished by a newly introduced VCP that allows us to attain the ratio- and CCC-based RDMs under any block segmentation. 
We prove that the IN terms 
in both RDMs 
approximately conform to centered Gaussian distributions whose variances are also derived. 
We further perform a comprehensive analysis comparing COS and the proposed sensing framework as well as the sensing performance under the two types of RDMs.
Extensive simulations validate the flexibility of the proposed sensing framework and its superiority over COS and \ccos.

\appendix

\subsection{Proof of Lemma \ref{lm: Sn[m] Wn[m] Zn[m] distributions}}\label{app: proof of lemma on S W Z distriutions}
As illustrated in Remark \ref{rmk: otfs represents dft-s-ofdm and ofdm}, $ s_n[l]\sim\mathcal{CN}(0,\sigma_d^2) $ and $ s_n[l]~(\forall n) $ is i.i.d. over $ l $.
As the unitary DFT of $ s_n[l] $, $ S_n[m]~(\forall n) $ conforms to the same distribution and is i.i.d. over $ m $.
Likewise, the distribution of $ W_n[m] $, which is the unitary DFT of $ w_n[l] $ given in (\ref{eq: w n[l]}), conforms to a complex Gaussian distribution and is i.i.d. over $ m $. Since $ w_n[l]~(\forall n) $ has the non-identical variance over $ l $, the variance of $ W_n[m] $ is not equal to that of $ w_n[l] $ and instead can be calculated, based on (\ref{eq: w n[l]}) and (\ref{eq: X n[m] new sensing}); specifically
\begin{align}
	{\big(2\sigma_w^2\tilde{Q} + \sigma_w^2(\tilde{M}-\tilde{Q})\big)}\Big/{\tilde{M}} = \left( 1 + {\tilde{Q}}/{\tilde{M}} \right)\sigma_w^2. \nonumber
\end{align}

Next, we analyze the distribution of $ Z_n[m] $. From (\ref{eq: X n[m] new sensing}), we see that the core of $ Z_n[m] $ is the DFT (w.r.t. $ l $) of $ z_n^{(p)}[l]g_{\tilde{Q}}[l]~(\forall p) $. 
From Fig. \ref{fig: rearrange and VCP}, we see that for any $ p $, $ z_n^{(p)}[l]g_{\tilde{Q}}[l] $ consists of two parts, one from the essential signal of the previous sub-block and the other from the sequential. Thus, $ z_n^{(p)}[l]g_{\tilde{Q}}[l]~(\forall p,~\forall n) $ satisfies $ z_n^{(p)}[l]g_{\tilde{Q}}[l]\sim\mathcal{CN}(0,\sigma_d^2) $, is i.i.d. over $ l $, and is independent from $ s_n[l] $.  
As a result, the DFT of $ z_n^{(p)}[l]g_{\tilde{Q}}[l] $ conforms to
\[\sum_{l=0}^{\tilde{M}-1} z_n^{(p)}[l]g_{\tilde{Q}}[l]  \myDFT{\tilde{M}}{lm}\sim\mathcal{CN}(0,\tilde{Q}\sigma_d^2\big/\tilde{M}).\]
With the assumption that $ {\alpha}_p $ is independent over $ p $, the $ p $-related summation in (\ref{eq: X n[m] new sensing}) leads to the final distribution of $ Z_n[m] $ given in the statement of the lemma.

\subsection{Proof of Lemma \ref{lm: Sn[m] Wn[m] Zn[m] dependence}} \label{app: proof of lemma on Sn[m] Wn[m] Zn[m] dependence}
The first statement of the lemma arises from the fact that $ W_n[m] $
is only related to the receiver noise while the other two components to the essential signals. Next, we illustrate the independence of $ S_n[m] $ over $ n $. 

Using the expression of $ S_n[m] $ given in (\ref{eq: X n[m] new sensing}),
we can have 
\begin{align}\label{eq: E{SS}}
	&\myExp{S_{n}[m]S_{n+1}^*[m]} = \mathbb{E}\left\{\left(\sum_{l_1=0}^{\tilde{M}-1} s_n[l_1]\myDFT{\tilde{M}}{l_1 m}\right)\times \right.\nonumber\\
	& \left.\left(\sum_{l_2=0}^{\tilde{M}-1} s_{n+1}^*[l_2]\myDFT{\tilde{M}}{-l_2 m}\right)\right\} = \bar{Q}\sigma_d^2 e^{-\mj \frac{2\pi(\tilde{M}-\bar{Q})}{\tilde{M}}}\Big/\tilde{M}.
\end{align}
The last result is based on two facts: \textit{First}, only at the indexes given in (\ref{eq: l1...l2=l1}), we can have non-zero expectation; otherwise the summands involved are uncorrelated. 
\begin{align} \label{eq: l1...l2=l1}
	l_1=\tilde{M}-\bar{Q},\cdots,\tilde{M}-1 \text{ and } l_2=l_1-(\tilde{M}-\bar{Q})
\end{align}
\textit{Second}, from Fig. \ref{fig: rearrange and VCP}, we can see that at the indexes given in (\ref{eq: l1...l2=l1}), $ s_n[l_1]=s_{n+1}[l_2] $. Based on (\ref{eq: E{SS}}), we can validate $ \myExp{|S_{n}[m]|^2}=\sigma_d^2 $ and $ \myExp{|S_{n+1}[m]|^2}=\sigma_d^2 $. Combining these expectations with (\ref{eq: E{SS}}), the correlation coefficient between $ S_{n}[m]$ and $S_{n+1}[m] $ can be obtained, as given in the statement of Lemma \ref{lm: Sn[m] Wn[m] Zn[m] dependence}. 
Following the above analysis procedure for $ S_n[m] $, we can similarly calculate the correlation coefficient between $ W_{n}[m]$ and $W_{n+1}[m] $. For brevity, we suppress the details here.

\subsection{Proof of Lemma \ref{lm: Zn[m] dependence over m}}
\label{app: proof of lemma on dependence Zn[m] over m}
As said in Appendix \ref{app: proof of lemma on S W Z distriutions}, $ S_n[m]~(\forall n) $ is the unitary DFT (w.r.t. $ l $) of $ s_n[l] $ that is i.i.d. over $ l $. Thus, $ S_n[m]~(\forall n) $ is i.i.d. over $ m $. For the same reason, $ W_n[m]~(\forall n) $ is i.i.d. over $ m $. However, $ Z_n[m] $
is not i.i.d. over $ m $, since the length of $ z_n^{(p)}[l]g_{\tilde{Q}}[l]~(\forall p) $ is smaller than the DFT dimension. Using the expression of $ Z_n[m] $ given in (\ref{eq: X n[m] new sensing}), 
we can calculate the correlation coefficient between $ Z_n[m_1] $ and $ Z_n[m_2] $. Specifically, we have 
\begin{align}\label{eq: E(Zn[m1]Zn[m2])}
	&\myExp{Z_n[m_1]Z_n^*[m_2]} \myEqualOverset{(a)} \sigma_P^2 \mathbb{E}\left\{\left(\sum_{l_1=0}^{\tilde{M}-1} z_n^{(p)}[l_1]g_{\tilde{Q}}[l_1]  \myDFT{\tilde{M}}{l_1m_1} \right)\right.\times\nonumber\\
	&	 
			\left.\left(\sum_{l_2=0}^{\tilde{M}-1} z_n^{(p)}[l_2]g_{\tilde{Q}}[l_2]  \myDFT{\tilde{M}}{l_2m_2}\right)^*\right\} 
	\myEqualOverset{(b)} \sigma_P^2\sigma_d^2  \frac{\sum_{l=0}^{\tilde{Q}-1} \myDFT{\tilde{M}}{l(m_1-m_2)}}{\sqrt{\tilde{M}}} ,
\end{align}
where 
$ \sigma_P^2 = \sum_{p=0}^{P-1}\sigma_p^2 $,
$ \myEqualOverset{(a)} $ is obtained based on the uncorrelated $ \alpha_p~(\forall p) $, and $ \myEqualOverset{(b)} $ is due to to the independence of $ z_n^{(p)}[l]g_{\tilde{Q}}[l] $ over $ l $.
Calculating the $ l $-related summation on the RHS of $ \myEqualOverset{(b)} $ by plugging in the definition of the DFT basis given in (\ref{eq: DFT basis}), we can obtain the following correlation coefficient
\begin{align}\label{eq: E{ZZ}/sqrt{E{Z}E{Z}}}
	&\frac{\left|\myExp{Z_n[m_1]Z_n^*[m_2]} \right|}{\sqrt{
			\myExp{|Z_n[m_1]|^2}
			\myExp{|Z_n[m_2]|^2}
		}
	} = \left|f(m_1-m_2)\right|, \\
	&\mathrm{s.t.}~f(m_1-m_2) = \substack{{\sin\left( \frac{2\pi}{\tilde{M}} \frac{\tilde{Q}(m_1-m_2)}{2} \right)}\Big/\left({\tilde{Q}\sin\left( \frac{2\pi}{\tilde{M}} \frac{(m_1-m_2)}{2} \right)}\right)} \nonumber
\end{align}
where $ \myExp{|Z_n[m_1]|^2} $ can be readily attained by setting $ m_2=m_1 $ in (\ref{eq: E(Zn[m1]Zn[m2])}); likewise for $ \myExp{|Z_n[m_2]|^2} $.

\subsection{Proof of Proposition \ref{pp: interference and noise of ratio-RDM}} \label{app: proof of proposition on IN of ratio-RDM}
As said above (\ref{eq: V k r l our rdm revised}), with a sufficiently large $ \ma $ introduced, $ {|\ma {S}_n[m]|}<1 $ can barely happen. Moreover, the operator $ \mathbb{I}_{\mathcal{E}}\{\cdot\} $ 
fully removes the cases of $ {|\ma {S}_n[m]|}<1 $. Therefore, we have
\begin{align}
	\left| {{D}_{n,m}^{k,l}}\big/{\ma {S}_n[m]} \right|\le \max\left\{ \left| \ma {D}_{n,m}^{k,l}  {S}_n^*[m] \right| \right\},
\end{align}
where $ {D}_{n,m}^{k,l} $ is defined in (\ref{eq: Z kr[l]+W kr[l] computation}). Since $ {{D}_{n,m}^{k,l}}\big/{\ma {S}_n[m]} $ conforms to a truncated Cauchy distribution, %
we are now able to invoke the CLT on deriving the distribution of $ \tilde{W}_k^{\mathrm{r}}[l] + \tilde{Z}_k^{\mathrm{r}}[l] $. 
Based on Lemma \ref{lm: Sn[m] Wn[m] Zn[m] dependence}, we know that $ Z_n[m] $ is independent over $ n $ and, under the condition $ \tilde{M}\gg (\tilde{Q}+\bar{Q}) $, such independence is also owned  by $ S_n[m] $ and $ W_n[m] $. Accordingly, the ratio $ {{D}_{n,m}^{k,l}}\big/{\ma {S}_n[m]}~(\forall m) $ is independent over $ n $. Invoking the CLT under large $ \tilde{N} $, we attain
\begin{align} \label{eq: Sum over n CN(0,)}
	&\sum_{n=0}^{\tilde{N}-1} \frac{{D}_{n,m}^{k,l}}{\ma {S}_n[m]}\sim\myCN{0,\tilde{N}\Big(\rho \mb(\epsilon)\Big)},\\
	&\mathrm{s.t.}~\rho =\frac{(\sigma_Z^2+\sigma_W^2)}{\tilde{M}\tilde{N}\ma^2\sigma_d^2},~b(\epsilon) = 2\left(  \ln\left( \frac{2(1-\epsilon)}{\sqrt{\epsilon(2-\epsilon)}} \right) - 1  \right),\nonumber
\end{align}
where $ \epsilon $ is a sufficiently small value and $ \rho\mb(\epsilon) $ is the variance of each summand according to \cite[Proposition 1]{Kai_otfs_IoTindustrial}. Note that $ \rho $ is the ratio between the variance of $ {D}_{2\tilde{n},m}^{k,l} $, as given in (\ref{eq: D nm kl CN(0,)}), and that of $ \ma {S}_{2\tilde{n}}[m] $, as easily deduced from Lemma \ref{lm: Sn[m] Wn[m] Zn[m] distributions}.

	With (\ref{eq: Sum over n CN(0,)})
attained, we know that $ \tilde{W}_k^{\mathrm{r}}[l] + \tilde{Z}_k^{\mathrm{r}}[l] $ also conforms to a Gaussian distribution, as it is the summation of $ \sum_{n=0}^{\tilde{N}-1} \frac{{D}_{n,m}^{k,l}}{\ma {S}_n[m]} $ over $ m $. 
Moreover, as shown below, $ \sum_{n=0}^{\tilde{N}-1} \frac{{D}_{n,m}^{k,l}}{\ma {S}_n[m]} $ is approximately independent over $ m $. This leads to the final distribution of $ \tilde{W}_k^{\mathrm{r}}[l] + \tilde{Z}_k^{\mathrm{r}}[l] $, as given in the statement of Proposition \ref{pp: interference and noise of ratio-RDM}.

\textit{Independence of $ \sum_{n=0}^{\tilde{N}-1} \frac{{D}_{n,m}^{k,l}}{\ma {S}_n[m]} $ over $ m $:} As Gaussian-distributed, $ \sum_{n=0}^{\tilde{N}-1} \frac{{D}_{n,m_1}^{k,l}}{\ma {S}_n[m_1]} $ and $ \sum_{n=0}^{\tilde{N}-1} \frac{{D}_{n,m_2}^{k,l}}{\ma {S}_n[m_2]} $ are independent if they are uncorrelated. 
Replacing $ {D}_{n,m}^{k,l} $ with its full expression given in (\ref{eq: Z kr[l]+W kr[l] computation}), we can have
\begin{align} \label{eq: E{Sum n1 Z/S Sum n2 Z/S}}
	& \myExp{\substack{\left(\sum_{n_1=0}^{\tilde{N}-1} \frac{ \left(Z_{n_1}[m_1] + {W}_{n_1}[m_1]\right) \myDFT{\bar{M}}{-m_1l}\myDFT{\tilde{N}}{n_1k} }{\ma {S}_{n_1}[m_1]}\right)\times \\
			\left(\sum_{n_2=0}^{\tilde{N}-1} \frac{ \left(Z_{n_2}[m_2] + {W}_{n_2}[m_2]\right) \myDFT{\bar{M}}{-m_2l}\myDFT{\tilde{N}}{n_2k} }{\ma {S}_{n_2}[m_2]}\right)^*}}\nonumber\\
	& \myEqualOverset{(a)} \myExp{\sum_{n=0}^{\tilde{N}-1} \frac{ {\substack{\left(Z_{n}[m_1] + {W}_{n}[m_1]\right)\left(Z_{n}^*[m_2] + {W}_{n}^*[m_2]\right)e^{\mj\frac{2\pi l(m_1-m_2)}{\tilde{M}}}}} }{{\ma^2S_n[m_1]S_n^*[m_2]\tilde{N}\tilde{M}} } }\nonumber\\
	& = \sum_{n=0}^{\tilde{N}-1} \mathbb{E}\left\{  f/g \right\} \overset{(b)}{\approx} \sum_{n=0}^{\tilde{N}-1}{\mu_f/\mu_g} =  0,\nonumber\\
	&\mathrm{s.t.}~f={\substack{
			\Big(Z_{n}[m_1]Z_{n}^*[m_2] +Z_{n}[m_1]{W}_{n}^*[m_2] +  {W}_{n}[m_1]Z_{n}^*[m_2]\\ +{W}_{n}[m_1]{W}_{n}^*[m_2] \Big)S_n^*[m_1]S_n[m_2]e^{\mj\frac{2\pi l(m_1-m_2)}{\tilde{M}}}}}\nonumber\\
	&~~~~~~ g=\ma^2|S_n[m_1]S_n[m_2]|^2\tilde{N}\tilde{M},
\end{align}%
where $ \myEqualOverset{(a)} $ is obtained by suppressing all cross-terms at $ n_1\ne n_2 $ since they are uncorrelated; and 
$ \overset{(b)}{\approx} $ is attained by applying the first-order approximation of the mean of the ratio$ f/g $ \cite{Note_seltman2012approximations}. The last result is based on $ \mu_f=0 $ which can be readily obtained by applying Lemmas \ref{lm: Sn[m] Wn[m] Zn[m] dependence} and \ref{lm: Zn[m] dependence over m}.

\subsection{Proof of Proposition \ref{pp: signal components in RDM ccc}} \label{app: proof of proposition on gaussian interfernce plus noise CCC}
Consider $ S_k^{\mathrm{c}}[l] $ first. According to Lemma \ref{lm: Sn[m] Wn[m] Zn[m] distributions}, we have $ S_n[m]\sim\mathcal{CN}(0,\sigma_d^2) $. This then yields $ \frac{|S_n[m]|^2}{\sigma_{{d}}^2/2}\sim\chi_2^2 $ and 
\begin{align}
	\myExp{ \frac{|S_n[m]|^2}{\sigma_{{d}}^2/2} }=2,~\myVar{ \frac{|S_n[m]|^2}{\sigma_{{d}}^2/2} }=4,
\end{align}
where $ \chi_2^2 $ denotes the chi-square distribution with two degrees of freedom (DoF).
Given that $ S_n[m]~(\forall n) $ is statistically independent of $ m $, we first consider the $ m $-related summation in calculating $ S_k^{\mathrm{c}}[l] $; see (\ref{eq: V k c l our rdm}). 
Invoking the CLT, we know that the summation leads to a Gaussian distribution whose variance is given by 
\begin{align}
	\myVar{\frac{\sigma_d^2}{2}
		{\sum_{m=0}^{\tilde{M}-1}
			\frac{|{S}_n[m]|^2}{\sigma_d^2/2}  e^{-\mj \frac{2\pi m \ml_p }{\tilde{M}}}  
			\myDFT{\tilde{M}}{-ml}}
	} = \sigma_d^4. \nonumber
\end{align} 
The mean of the Gaussian distribution is more complicated, as we need to consider the cases of $ l=\ml_p $ and $ \l\ne \ml_p $. 
For the first case, we have 
\begin{align}
	\myExp{\frac{\sigma_d^2}{2}
		{\sum_{m=0}^{\tilde{M}-1}
			\frac{|{S}_n[m]|^2}{\sigma_d^2/2}  e^{-\mj \frac{2\pi m \ml_p }{\tilde{M}}}  
			\myDFT{\tilde{M}}{-m\ml_p}}} = \sigma_d^2\sqrt{\tilde{M}},\nonumber
\end{align}
where $ e^{-\mj \frac{2\pi m \ml_p }{\tilde{M}}}  
\myDFT{\tilde{M}}{-m\ml_p} = 1/\sqrt{\tilde{M}} $ is applied. For the case of $ l\ne \ml_p $, we have 
\begin{align}
	\myExp{\frac{\sigma_d^2}{2}
		{\sum_{m=0}^{\tilde{M}-1}
			\frac{|{S}_n[m]|^2}{\sigma_d^2/2}  e^{-\mj \frac{2\pi m \ml_p }{\tilde{M}}}  
			\myDFT{\tilde{M}}{-ml}}} &= \sigma_d^2 \sum_{m=0}^{\tilde{M}-1}
	e^{\mj \frac{2\pi m (l-\ml_p) }{\tilde{M}}} \nonumber\\
	&=0,
\end{align}
where we have applied the fact that summing a discrete single-tone exponential signal (with a non-zero frequency) over integer cycles leads to zero. Combining the above discussion, we conclude  
\begin{align} %
	\sum_{m=0}^{\tilde{M}-1}
	{|{S}_n[m]|^2}  e^{-\mj \frac{2\pi m \ml_p }{\tilde{M}}}  
	\myDFT{\tilde{M}}{-ml}\sim\left\{
	\begin{array}{ll}
		\myN{\sigma_d^2\sqrt{\tilde{M}},\sigma_d^4} & l= \ml_p\\
		\myN{0,\sigma_d^4} & l\ne \ml_p
	\end{array}
	\right.. \nonumber
\end{align}
Based on Lemma \ref{lm: Sn[m] Wn[m] Zn[m] dependence}, $ \sum_{m=0}^{\tilde{M}-1}
{|{S}_n[m]|^2}  e^{-\mj \frac{2\pi m \ml_p }{\tilde{M}}}  
\myDFT{\tilde{M}}{-ml} $ is independent over the set of either odd $ n $ or even $ n $, but shows dependence between adjacent pair. 
However, under the assumption of $ \tilde{M}\gg (\tilde{Q}+\bar{Q}) $, the dependence is weak as the correlation coefficient approaches zero. Thus, summing the LHS of the above equation over $ n $ converges in distribution to another Gaussian with its statistical properties depicted in the statement of Proposition \ref{pp: signal components in RDM ccc}.
The distribution of $ \mX_k^{\mathrm{r}}[l],~\mX\in\{W,Z\} $ can be similarly derived with reference to Appendix \ref{app: proof of proposition on IN of ratio-RDM}. 
Thus, we suppress the details for brevity.

\bibliographystyle{IEEEtran}
\bibliography{IEEEabrv,./ref/bib_JCAS.bib}

\begin{thebibliography}{10}
\providecommand{\url}[1]{#1}
\csname url@samestyle\endcsname
\providecommand{\newblock}{\relax}
\providecommand{\bibinfo}[2]{#2}
\providecommand{\BIBentrySTDinterwordspacing}{\spaceskip=0pt\relax}
\providecommand{\BIBentryALTinterwordstretchfactor}{4}
\providecommand{\BIBentryALTinterwordspacing}{\spaceskip=\fontdimen2\font plus
\BIBentryALTinterwordstretchfactor\fontdimen3\font minus
  \fontdimen4\font\relax}
\providecommand{\BIBforeignlanguage}[2]{{%
\expandafter\ifx\csname l@#1\endcsname\relax
\typeout{** WARNING: IEEEtran.bst: No hyphenation pattern has been}%
\typeout{** loaded for the language `#1'. Using the pattern for}%
\typeout{** the default language instead.}%
\else
\language=\csname l@#1\endcsname
\fi
#2}}
\providecommand{\BIBdecl}{\relax}
\BIBdecl

\bibitem{FanLiu_overview2020TCOM}
F.~{Liu}, C.~{Masouros}, A.~P. {Petropulu}, H.~{Griffiths}, and L.~{Hanzo},
  ``Joint radar and communication design: Applications, state-of-the-art, and
  the road ahead,'' \emph{IEEE Trans. Commun.}, vol.~68, no.~6, pp. 3834--3862,
  2020.

\bibitem{DFRC_SP_Mag2019Amin_Aboutanios}
A.~{Hassanien}, M.~G. {Amin}, E.~{Aboutanios}, and B.~{Himed}, ``Dual-function
  radar communication systems: A solution to the spectrum congestion problem,''
  \emph{IEEE Signal Process. Mag.}, vol.~36, no.~5, pp. 115--126, Sep. 2019.

\bibitem{DFRC_automotive2020SPmag}
D.~Ma, N.~Shlezinger, T.~Huang, Y.~Liu, and Y.~C. Eldar, ``Joint
  radar-communication strategies for autonomous vehicles: Combining two key
  automotive technologies,'' \emph{IEEE Signal Process. Mag.}, vol.~37, no.~4,
  pp. 85--97, 2020.

\bibitem{DFRC_radarCentric_mishra2019toward}
K.~V. Mishra, M.~B. Shankar, V.~Koivunen, B.~Ottersten, and S.~A. Vorobyov,
  ``Toward millimeter-wave joint radar communications: A signal processing
  perspective,'' \emph{IEEE Signal Process. Mag.}, vol.~36, no.~5, pp.
  100--114, 2019.

\bibitem{Kai_overviewFHMIMO_DFRC2020AES}
K.~Wu, J.~A. Zhang, X.~Huang, and Y.~J. Guo, ``Frequency-hopping {MIMO}
  radar-based communications: An overview,'' \emph{arXiv preprint
  arXiv:2006.07559}, 2020.

\bibitem{DFRC_802p11ad2018TVT_Kumari}
P.~Kumari, J.~Choi, N.~González-Prelcic, and R.~W. Heath, ``{IEEE}
  802.11ad-based radar: An approach to joint vehicular communication-radar
  system,'' \emph{IEEE Trans. Vehic. Techn.}, pp. 67(4) 3012--3027, 2018.

\bibitem{DFRC_OpportunisticRadar_80211ad_2018TSP}
E.~{Grossi}, M.~{Lops}, L.~{Venturino}, and A.~{Zappone}, ``Opportunistic radar
  in {IEEE} 802.11ad networks,'' \emph{IEEE Trans. Signal Process.}, vol.~66,
  no.~9, pp. 2441--2454, May 2018.

\bibitem{DFRC_CommCentric_duggal2020doppler}
G.~Duggal, S.~Vishwakarma, K.~V. Mishra, and S.~S. Ram, ``Doppler-resilient
  802.11 ad-based ultrashort range automotive joint radar-communications
  system,'' \emph{IEEE Trans. Aerospace Electr. Syst.}, vol.~56, no.~5, pp.
  4035--4048, 2020.

\bibitem{DFRC_QixunZhang9162963}
Q.~Zhang, H.~Sun, Z.~Wei, and Z.~Feng, ``Sensing and communication integrated
  system for autonomous driving vehicles,'' in \emph{IEEE INFOCOM 2020 - IEEE
  Conference on Computer Communications Workshops (INFOCOM WKSHPS)}, 2020, pp.
  1278--1279.

\bibitem{DFRC_dsss2011procIeee}
C.~{Sturm} and W.~{Wiesbeck}, ``Waveform design and signal processing aspects
  for fusion of wireless communications and radar sensing,'' \emph{Proc. IEEE},
  vol.~99, no.~7, pp. 1236--1259, 2011.

\bibitem{DFRC_SC_OFDM}
Y.~Zeng, Y.~Ma, and S.~Sun, ``Joint radar-communication with cyclic prefixed
  single carrier waveforms,'' \emph{IEEE Trans. Veh. Techn.}, vol.~69, no.~4,
  pp. 4069--4079, 2020.

\bibitem{DFRC_JointDesign_9354629}
T.~Wild, V.~Braun, and H.~Viswanathan, ``Joint design of communication and
  sensing for beyond {5G} and {6G} systems,'' \emph{IEEE Access}, vol.~9, pp.
  30\,845--30\,857, 2021.

\bibitem{DFRC_JointDesign_9148935}
C.~B. Barneto, S.~D. Liyanaarachchi, T.~Riihonen, L.~Anttila, and M.~Valkama,
  ``Multibeam design for joint communication and sensing in {5G} new radio
  networks,'' in \emph{2020 IEEE International Conference on Communications
  (ICC)}, 2020, pp. 1--6.

\bibitem{DFRC_AngLi_liu2021cram}
F.~Liu, Y.-F. Liu, A.~Li, C.~Masouros, and Y.~C. Eldar, ``Cram\'er-rao bound
  optimization for joint radar-communication design,'' \emph{arXiv preprint
  arXiv:2101.12530}, 2021.

\bibitem{Andrew_jcasOverview2021JSTSP}
J.~A. Zhang, F.~Liu, C.~Masouros, R.~W. Heath~Jr, Z.~Feng, L.~Zheng, and
  A.~Petropulu, ``An overview of signal processing techniques for joint
  communication and radar sensing,'' \emph{arXiv:2102.12780}, 2021.

\bibitem{Kai_rahman2020enablingSurvey}
M.~L. Rahman, J.~A. Zhang, K.~Wu, X.~Huang, Y.~J. Guo, S.~Chen, and J.~Yuan,
  ``Enabling joint communication and radio sensing in mobile networks--a
  survey,'' \emph{arXiv preprint arXiv:2006.07559}, 2020.

\bibitem{DFRC_wei2021towards}
Z.~Wei, F.~Liu, C.~Masouros, N.~Su, and A.~P. Petropulu, ``Towards
  multi-functional {6G} wireless networks: Integrating sensing, communication
  and security,'' \emph{arXiv preprint arXiv:2107.07735}, 2021.

\bibitem{DFRC_4IoT_cui2021integrating}
Y.~Cui, F.~Liu, X.~Jing, and J.~Mu, ``Integrating sensing and communications
  for ubiquitous {IoT}: Applications, trends and challenges,'' \emph{arXiv
  preprint arXiv:2104.11457}, 2021.

\bibitem{OFDM_autonomousDriv2019microwaveMag}
F.~{Roos}, J.~{Bechter}, C.~{Knill}, B.~{Schweizer}, and C.~{Waldschmidt},
  ``Radar sensors for autonomous driving: Modulation schemes and interference
  mitigation,'' \emph{IEEE Microw. Mag.}, vol.~20, no.~9, pp. 58--72, 2019.

\bibitem{BinYang_ofdmSPmagazine}
G.~{Hakobyan} and B.~{Yang}, ``High-performance automotive radar: A review of
  signal processing algorithms and modulation schemes,'' \emph{IEEE Signal
  Process. Mag.}, vol.~36, no.~5, pp. 32--44, 2019.

\bibitem{OTFS_Magazine_wei2020orthogonal}
Z.~Wei, W.~Yuan, S.~Li, J.~Yuan, G.~Bharatula, R.~Hadani, and L.~Hanzo,
  ``Orthogonal time-frequency space modulation: A full-diversity next
  generation waveform,'' \emph{arXiv preprint arXiv:2010.03344}, 2020.

\bibitem{OTFS_jcas2020twc}
L.~Gaudio, M.~Kobayashi, G.~Caire, and G.~Colavolpe, ``On the effectiveness of
  {OTFS} for joint radar parameter estimation and communication,'' \emph{IEEE
  Trans. Wireless Commun.}, vol.~19, no.~9, pp. 5951--5965, 2020.

\bibitem{OTFS_yuan2021integratedSensingCOmmunicationOTFS}
W.~Yuan, Z.~Wei, S.~Li, J.~Yuan, and D.~W.~K. Ng, ``Integrated sensing and
  communication-assisted orthogonal time frequency space transmission for
  vehicular networks,'' \emph{arXiv:2105.03125}, 2021.

\bibitem{OTFS_Raviteja2019TVT_embeddedPilotChannelEstimation}
P.~{Raviteja}, K.~T. {Phan}, and Y.~{Hong}, ``Embedded pilot-aided channel
  estimation for {OTFS} in delay–doppler channels,'' \emph{IEEE Trans. Veh.
  Techn.}, vol.~68, no.~5, pp. 4906--4917, 2019.

\bibitem{OTFS_keskin2021radarTimeDomainICIisi}
M.~F. Keskin, H.~Wymeersch, and A.~Alvarado, ``Radar sensing with {OTFS}:
  Embracing {ISI} and {ICI} to surpass the ambiguity barrier,'' \emph{arXiv
  preprint arXiv:2103.16162}, 2021.

\bibitem{OTFS_windowDesignWeizhiQiang2021}
Z.~Wei, W.~Yuan, S.~Li, J.~Yuan, and D.~W.~K. Ng, ``Transmitter and receiver
  window designs for orthogonal time-frequency space modulation,'' \emph{IEEE
  Trans. Commun.}, vol.~69, no.~4, pp. 2207--2223, 2021.

\bibitem{OTFS_DasOFDMbasedOTFSIEEEaccess2021}
S.~S. Das, V.~Rangamgari, S.~Tiwari, and S.~C. Mondal, ``Time domain channel
  estimation and equalization of {CP-OTFS} under multiple fractional dopplers
  and residual synchronization errors,'' \emph{IEEE Access}, vol.~9, pp.
  10\,561--10\,576, 2021.

\bibitem{OTFS_Raviteja2019TVTpulseShapingRCP}
P.~Raviteja, Y.~Hong, E.~Viterbo, and E.~Biglieri, ``Practical pulse-shaping
  waveforms for reduced-cyclic-prefix {OTFS},'' \emph{IEEE Trans. Veh. Techn.},
  vol.~68, no.~1, pp. 957--961, 2019.

\bibitem{Gaussian_OFDMenvelope}
S.~{Wei}, D.~L. {Goeckel}, and P.~A. {Kelly}, ``Convergence of the complex
  envelope of bandlimited {OFDM} signals,'' \emph{IEEE Trans. Information
  Theory}, vol.~56, no.~10, pp. 4893--4904, 2010.

\bibitem{book_oppenheim1999discrete}
A.~V. Oppenheim, \emph{Discrete-time signal processing}.\hskip 1em plus 0.5em
  minus 0.4em\relax Pearson Education India, 1999.

\bibitem{book_richards2010principlesModernRadar}
M.~A. Richards, J.~Scheer, W.~A. Holm, and W.~L. Melvin, \emph{Principles of
  modern radar}.\hskip 1em plus 0.5em minus 0.4em\relax Citeseer, 2010.

\bibitem{book_ahmadi2019_5G}
S.~Ahmadi, \emph{{5G} {NR}: Architecture, Technology, Implementation, and
  Operation of {3GPP} New Radio Standards}.\hskip 1em plus 0.5em minus
  0.4em\relax Academic Press, 2019.

\bibitem{book_van2004optimum}
H.~L. Van~Trees, \emph{Optimum array processing: Part IV of detection,
  estimation, and modulation theory}.\hskip 1em plus 0.5em minus 0.4em\relax
  John Wiley \& Sons, 2004.

\bibitem{Kai_padeFreqEst2021TVT}
K.~Wu, J.~A. Zhang, X.~Huang, and Y.~J. Guo, ``Accurate frequency estimation
  with fewer {DFT} interpolations based on {Pad\'e} approximation,''
  \emph{arXiv preprint arXiv:2105.13567}, 2021.

\bibitem{Kai_ofdmSensingSPM}
------, ``A low-complexity method for {FFT}-based {OFDM} sensing,'' \emph{arXiv
  preprint arXiv:2105.13596}, 2021.

\bibitem{CFAR_FFTimplement_kronauge2013fast}
M.~Kronauge and H.~Rohling, ``Fast two-dimensional cfar procedure,'' \emph{IEEE
  Trans. Aerospace Electr. Syst.}, vol.~49, no.~3, pp. 1817--1823, 2013.

\bibitem{ComplexGaussianRatio_2010Globecom}
R.~J. {Baxley}, B.~T. {Walkenhorst}, and G.~{Acosta-Marum}, ``Complex gaussian
  {Ratio} distribution with applications for error rate calculation in fading
  channels with imperfect csi,'' in \emph{2010 IEEE Global Telecommunications
  Conference GLOBECOM 2010}, 2010, pp. 1--5.

\bibitem{Cauchy_truncated_hampel1998statistics}
F.~Hampel and E.~Zurich, ``Is statistics too difficult?'' \emph{Canadian
  Journal of Statistics}, vol.~26, no.~3, pp. 497--513, 1998.

\bibitem{Kai_otfs_IoTindustrial}
K.~Wu, J.~A. Zhang, X.~Huang, and Y.~J. Guo, ``{OTFS}-based joint communication
  and sensing for future industrial {IoT},'' \emph{arXiv preprint}, 2021.

\bibitem{Gaussian_productOfcomplexGaussians}
N.~O'Donoughue and J.~M.~F. Moura, ``On the product of independent complex
  {Gaussians},'' \emph{IEEE Trans. Signal Process.}, vol.~60, no.~3, pp.
  1050--1063, 2012.

\bibitem{Note_seltman2012approximations}
H.~Seltman, ``Approximations for mean and variance of a ratio,''
  \emph{unpublished note}, 2012.

\end{thebibliography}
\end{document}